\documentclass[useAMS,usenatbib]{mn2e}
\usepackage{times}
\usepackage{epsfig}
\usepackage{graphicx}
\usepackage[usenames,dvips]{color}

\title[Binaries with NS in globular clusters]{Formation and evolution of compact binaries in globular clusters: 
II.~Binaries with neutron stars.}
\author[N Ivanova et al.]
{N.\ Ivanova $^1$\thanks{E-mail:nata@cita.utoronto.ca, Tremaine Fellow}, C.~0.\ Heinke$^{2,3}$\thanks{Lindheimer Fellow}, F.~A.\ Rasio$^2$, 
K.\ Belczynski$^{4,5}$\thanks{Oppenheimer Fellow},
\& J.~M.\ Fregeau$^{2}$\thanks{Chandra Fellow} \\
$^1$Canadian Institute for Theoretical Astrophysics, University of Toronto, 60 St. George, Toronto, ON M5S 3H8, Canada\\
$^2$Northwestern University, Dept. of Physics \& Astronomy,  2145 Sheridan Rd, Evanston, IL 60208, USA\\
$^3$University of Virginia, Dept. of Astronomy, PO Box 400325, Charlottesville, VA 22904, USA \\
$^4$Los Alamos National Lab,    P.O. Box 1663, MS 466, Los Alamos, NM 87545 \\
$^5$New Mexico State University, Department of Astronomy,  1320 Frenger Mall, Las Cruces, New Mexico 88003-8001, USA
}

\begin{document}

\maketitle

\label{firstpage}

\begin{abstract}{
In this paper, the second of a series, 
we study the stellar dynamical  and evolutionary
processes leading to the formation of
compact binaries containing neutron stars (NSs) in dense globular clusters. 
For this study, 70 dense clusters were simulated independently, with a
total stellar mass $\sim 2\times10^7\ M_\odot$,
exceeding the total mass of all dense globular clusters in our Galaxy.

We find that, in order to reproduce the empirically derived formation rate of 
low-mass X-ray binaries (LMXBs),
we must assume that NSs can be formed via electron-capture supernovae 
with typical natal kicks smaller than in core-collapse supernovae.
Our results explain the observed dependence of the number of LMXBs on ``collision number''
as well as the large scatter observed between different globular clusters.
We predict that the number of quiescent LMXBs in different clusters should not have a strong metallicity dependence.

We compare the results obtained from our simulations with the observed population of 
millisecond pulsars (MSPs).  
We find that in our cluster model the following mass-gaining events create populations of MSPs that 
do not match the observations (either they are inconsistent with the observed LMXB production rates, or the inferred 
binary periods or companion masses are not observed among radio bMSPs): (i) accretion during a common envelope event with a
NS formed through electron capture supernovae, and (ii) mass transfer (MT) from a WD donor. 
Some processes lead only to a mild recycling -- physical collisions or MT in a post-accretion induced collapse system.
In addition, for MSPs, we distinguish low-magnetic-field (long-lived) and 
high-magnetic-field (short-lived) populations, where in the latter NSs are formed as a result of
accretion induced collapse or merger induced collapse.
 With this distinction and by considering only those mass-gaining events that appear to lead to NS recycling, 
we obtain good agreement of our models with the 
numbers and characteristics of observed MSPs 
in 47~Tuc and Terzan 5,  as well as with the 
cumulative statistics for MSPs detected in globular clusters of different dynamical properties. 

We find that significant production of 
merging double NSs potentially detectable as 
short $\gamma$-ray bursts occurs only in very dense, most likely core-collapsed clusters.
}
\end{abstract}


\section{Introduction}

Neutron stars (NSs) are seen in globular clusters (GCs) 
via their (current or past) participation in interacting binary systems.
From the earliest observations of X-ray binaries in GCs
it has been noted that their abundance per unit mass is $\sim$100 times  
greater in GCs than in the Galaxy as a whole \citep{Katz75}.
This was understood to be a consequence of the high stellar density of GCs, 
which may lead to the creation of compact NS binaries in close stellar encounters \citep{Clark75}.
NS binaries are often relatively easy to identify in GCs (compared to other binaries), 
and they allow us to constrain the dynamical history of clusters and 
the evolution of NS properties in binary systems.  
They have been observed so far in three guises:
(i) as bright low-mass X-ray binaries (LMXBs); 
(ii) as quiescent LMXBs (qLMXBs); and (iii) as binary or single millisecond pulsars (MSPs).
In the future, they could also be detected as sources of gravitational radiation.

Bright LMXBs have X-ray luminosities of typically $10^{35}$-- $10^{38}$ ergs s$^{-1}$.  
In bright LMXBs, the NS is 
actively accreting from a companion star via Roche-lobe overflow and an accretion disk.  
Thirteen sources are known in Galactic GCs, of which 12 are known to contain NSs 
through the detection of thermonuclear bursts.  Six are transient systems (two have 
remained in outburst for more than a decade; the others have outburst time-scales of 
weeks), while the other seven seem to be persistent.  Orbital periods have now been identified for seven 
systems \citep{Verbunt06,Dieball05,Altamirano07}, of which three are below one hour, indicating ultracompact 
systems composed of NSs with white dwarf companions are common in globular clusters \citep{Deutsch00_ucxb}. 
Efficient surveys of the X-ray sky give us confidence that we have detected all 
bright LMXBs that have existed in Galactic GCs during the past 10 years. 

Quiescent LMXBs have lower X-ray luminosities 
($10^{31}<L_X<10^{34}$ ergs/s), indicating that accretion is not active or is 
significantly reduced during quiescence.  
Their X-ray luminosity is thought to derive at least in part from the re-radiation of heat from 
the core of the NS, which was generated during accretion outbursts \citep{Brown98}, 
though other mechanisms are also suggested (see, e.g., \cite{Campana98a} ).
They can be identified through their  positional coincidence with the location of 
a bright transient LMXB in outburst \citep{Wijnands05}, and/or through their characteristically 
soft X-ray spectrum \citep{2002ApJ...578..405R}. \footnote{Not all qLMXBs show the 
characteristic soft X-ray spectrum \citep{Wijnands05}, though it seems that the 
majority do.}  The total number of qLMXBs in the Galactic GC system is 
inferred to be $\sim$100-200 \citep{Heinke03a}, although only a fraction of the Galactic 
clusters have been studied sufficiently to identify qLMXBs.  

As of this writing 129 millisecond radio pulsars have been observed in GCs,
with more than 10, 20 and 30 seen from the clusters 
M28 \citep{Stairs06}, 47~Tuc \citep{Camilo00}, and Terzan~5 \citep{Ransom05}, 
respectively\footnote{See http://www.naic.edu/~pfreire/GCpsr.html for an updated list.}.  
These millisecond pulsars (MSPs) are thought to be descendants of  
LMXBs \citep[for reviews see][]{Bhattacharya91,Phinney94}.  
It is estimated that of order 1000 MSPs are present in the Galactic GC system 
\citep{Heinke05a}, of which half or more are potentially detectable.
Current pulsar searches have reached the bottom of the pulsar luminosity function in very few clusters (e.g., 47~Tuc and M15).
As an example, the total radio luminosity from Terzan~5 suggested that this 
cluster may contain 100 pulsars \citep{Fruchter95_ter5}---this number is 3 times more than the number of identified pulsars.  

Since NSs in GCs appear to be excellent tests for our understanding of both
stellar dynamics and neutron star evolution, it is not surprising that
this issue has attracted significant continuing attention.
The largest challenge for these studies is computational.
It has been estimated that a cluster consisting of $10^6$ stars ($\sim 250,000 M_\odot$) will contain
only a few hundred NSs, and models with a smaller number of stars will
not therefore have sufficient statistics. 
Direct $N$-body methods are unlikely to address the million-body problem before 2020 \citep{Hut_nbody}.
Monte Carlo methods now allow million-body simulations for spherically symmetric systems 
with single stars \citep{Giersz06} and with large fractions of
primordial binaries, including direct integrations of all binary encounters  \citep{Fregeau06},
but a full implementation of binary star evolution is still lacking \citep{Rasio06}.
As a result, most previous theoretical studies of NSs in clusters 
were performed using analytic esimates \citep[e.g.,][]{Hut91},
or no binaries \citep[e.g.,][]{Drukier96}, or
simple scattering experiments, with a static background of single or binary stars
 \citep[e.g.,][]{1995ApJS...99..609S, Rasio00,  Kuranov06}. 
We chose instead to use a method that treats in detail the stellar evolution of a large population 
of stars and binaries in a dense cluster environment, but with a highly simplified description 
of the stellar dynamics which is not fully self-consistent.

In this paper we present the results of our ongoing study of compact binary formation in GCs, 
concentrating on the formation of binaries with NSs. 
For the first time we have used cluster simulations that include a dynamically evolving
system and represent consistently an entire 
stellar population with numbers of stars and binaries comparable to those  in observed GCs. 
In addition, several statistically independent simulations were performed  
for each cluster model, to show how different 
the outcomes can be with slightly different but equivalent initial conditions. 
Some preliminary results on NS binaries from our work 
were reported in \citet{Ivanova04_proc_bgc,Ivanova05_proc_nsgc,Ivanova05_proc_wdgc,
Rasio06}, and detailed results for the formation of binaries 
containing white dwarfs (WDs) were presented in  Paper~I \citep{Ivanova06_wd}.
We refer the reader to Paper~I for a complete description of the physical processes
of formation and destruction of mass-transferring compact binaries, as
well as our numerical methods.
We outline updates to the method and our assumptions that are especially
relevant to NSs in Section ~2.
Section 3 outlines the scenarios for formation of MSPs in GCs.
In Section~4 we describe the range of GC models that we consider.
Formation and retention of NSs are analyzed in Section~5, and the following
section is devoted to the formation of binaries containing NSs.
We discuss the formation of LMXBs in Section~7, and MSPs
are considered in Section~8. Section~9 briefly describes double neutron stars. 
We conclude by summarizing the connection between our
results and the observations, as well as giving predictions for future observations
and discussing constraints on NS formation.

\section{Method}

Our numerical methods generally remain the same as
described in \cite{Ivanova05_bf,Ivanova06_wd}. 
To study the population of NSs, we introduced 
a number of updates to the population synthesis model
specifically for NS formation,  and we revised the treatment 
of dynamical events with NSs. We also optimized numerical
aspects of our code for more efficient use of parallel supercomputers.

\subsection{Population synthesis updates}

We adopt the binary evolution model from the population synthesis 
code {\tt StarTrack} \citep{Bel02,Bel06_startrack}. 
In our first study of binary populations in GC cores \citep{Ivanova05_bf} 
we used a previous version of {\tt StarTrack} described in \cite{Bel02}.
In our study of close binaries with  white dwarfs \citep[][paper I]{Ivanova06_wd} 
we employed a more recent version of {\tt StarTrack} available at the time
\citep[as in][ but including updates only for WD evolution compared to the previous version]{Bel06_startrack}.
In this paper we incorporated all the latest updates of {\tt StarTrack}, as  
 described in \citet{Bel06_startrack}.
Below we outline several of the most important changes in {\tt StarTrack}
that affect the formation and evolution of NSs compared to our previous studies of NSs in GCs 
\citep{Ivanova04_proc_bgc,Ivanova05_proc_nsgc,Ivanova06_wd}.
We also provide detailed descriptions of further modifications to the 
{\tt StarTrack} model, beyond those in  \citet{Bel06_startrack}, which we developed
for the treatments  of electron-capture supernovae 
and common-envelope events.

\subsubsection{Natal kicks and supernova disruptions}

At the moment of formation, both NSs and BHs receive 
 additional speed (a natal kick), most likely due 
to asymmetric supernova ejecta \citep{Fryer04_kicks}.
Although most recent simulations are in relatively good agreement with the
measured distribution 
of pulsar velocities, the agreement is not yet firmly established \citep{Scheck06_kicks}.
In this paper we adopt the most recently derived pulsar kick velocity distribution 
from \cite{Hobbs05_kicks}, which is 
a Maxwellian distribution with one-dimensional RMS velocity
$\sigma = 265$ km s$^{-1}$  (the mean three-dimensional velocity is $\sim 400 $ km s$^{-1}$).
In contrast to some earlier studies \citep[e.g., see][]{Arzoumanian02ApJ_kicks},
no evidence for a bimodal velocity distribution is present.

At the moment of the explosion, the supernova progenitor 
can be a binary companion. Because of the kick, the binary can become unbound 
(and the binary components will separate) or the entire binary can be 
affected by the natal kick. In our previous studies, in order to find the 
velocities of the components of an unbound binary, 
we used the derivation by \cite{Tauris98_kicks},
where the pre-supernova orbit is assumed to be circular.
The approach used in the current version of {\tt StarTrack}
allows us to properly calculate cases with an arbitrary 
pre-explosion orbital eccentricity \citep[for full details, see \S 6.3 in][]{Bel06_startrack}.

\subsubsection{Electron-capture supernovae}

Through our studies of accreting WDs \citep{Ivanova06_wd} we found that 
electron-capture supernovae (ECSe) could be the dominant source 
of retained NSs in clusters.
Similar results were also recently obtained in \citet{Kuranov06}.
We have therefore considered in great detail all possible types of ECSe for this work.

It has been argued that when a {\em degenerate\/} ONeMg core reaches  $M_{\rm ecs}=1.38 M_\odot$,
its collapse is triggered by  electron capture on $^{24}$Mg and $^{20}$Ne 
before neon and subsequent burnings start and, therefore, before the iron core formation 
\citep{Miyaji80_ecs, Nomoto84_ecs1, Nomoto87_ecs2,Timmes92_cotoonems,Timmes94_cotoonems}. 
The explosion energy of such an event is 
significantly lower than that inferred for core-collapse supernovae \citep{Dessart06_aic,Kitaura06}.
There are several possible situations when a star can develop a degenerate ONeMg core
that will eventually reach $M_{\rm ecs}$.

First, this can occur during the normal evolution of single stars.
It was initially proposed by \cite{Barkat74_ironcore} that a star of 7--10 $M_\odot$,
after non-explosive carbon burning, develops an ONeMg core.
If the initial core mass is less than required for the neon ignition, $1.37 M_\odot$,
the core becomes strongly degenerate. Through the continuing He shell burning, this
core grows to $M_{\rm ecs}$. 
In more massive stars, $\ga 10\ M_\odot$, carbon, oxygen, neon and silicon burnings
progress under non-degenerate conditions, and, in less massive stars, ONeMg cores never form.
The formation of degenerate or non-degenerate ONeMg cores depends on
the He core mass at the start of the asymptotic giant branch, where
 $\sim 2.5\ M_\odot$ is a rough boundary
between the cases \citep{Nomoto84_ecs1}. 
If the initial He core mass is below $1.83\ M_\odot$, no off-centre ignition will happen---the 
carbon core burning will occur when the degenerate core reaches the Chandrasekhar mass,
resulting in a thermonuclear explosion \cite[][also J.~Eldridge 2006, priv. comm.]{Hurley00}.
 
The conditions for ECS were shown to occur in single stars of initial mass 
8--10 $M_\odot$ by \citet{Nomoto84_ecs1}.
This critical mass range depends on the properties of the He and CO cores, which, in turn, 
are highly dependent on the mixing prescription (semiconvection, overshooting, rotational mixing, etc.)
as well as on the adopted opacities. As a result, the range varies between different evolutionary codes 
\citep[see discussion in][]{Podsi04_aic,Siess06_agb},
and is generally reduced compared to that proposed initially by \cite{Nomoto84_ecs1,Nomoto87_ecs2}.
In the code that we use for our cluster simulations, a non-degenerate ONeMg core
is formed when the initial He core mass is about $2.25\ M_\odot$ \citep{Pols98_models,Hurley00} and 
the range of initial masses for single stars of solar metallicity that leads to the formation
of such a core is  7.66 to 8.26 $M_\odot$.
The equivalent mass ranges are from 6.85 to 7.57  $M_\odot$ and from  6.17 to 6.76  $M_\odot$
for single stars with GC-like metallicities, $Z=0.005$ and  $Z=0.0005$, respectively. 

The initial stellar mass is not the only parameter that defines whether a star will form a 
degenerate ONeMg core. 
It was recently pointed out by  \cite{Podsi04_aic} that the range of progenitor masses
for which an ECS can occur depends also on the
mass transfer history of the star, and therefore may be different for 
binary stars, making it possible for more massive progenitors to collapse via ECS. 
We will refer to the single and binary scenarios for ECS in
 non-degenerate stars, as described above, as evolution-induced collapse (EIC).

The second possibility for an ECS to occur is by accretion on to a 
degenerate ONeMg WD in a binary: accretion-induced collapse (AIC).
In this case, a massive ONeMg WD steadily accumulates mass until it reaches 
the critical mass $M_{\rm ecs}$ 
\citep[for more detail on the adopted model of the accretion on the WD see][]
{Ivanova04_ttmt,Bel06_startrack}.
In addition, ONeMg WDs can be formed from CO WDs via off-centre carbon ignition if the mass of CO WDs
is above 1.07 $M_\odot$ and the accretion rate is high,
$\dot M > 2.7 \times 10^{-6} M_\odot {\rm yr^{-1}}$ \citep{Kawai87_co} so that eventually this
also leads to AIC. 

The third case that we consider is coalescing double WDs with a total mass
exceeding $M_{\rm ecs}$. To distinguish it from AIC, we will refer to this case
as merger-induced collapse (MIC). The nature of NS formation is the same
as in the case of AIC (accumulation of mass by a massive ONeMG WD).
In the case of coalescing CO WDs, then, as in the case of single stars,
off-centre carbon ignition occurs first, which quiescently converts the star into an ONeMg core,
and this proceeds with an ECS \citep{Saio85_cowd,Saio04_cowd}.
Although the approach above was proposed for coalescing WDs in binaries only,
we assume that in the case of {\em collisions\/} between two WDs it is applicable as well: except 
for the rare cases of near head-on collisions, a less massive WD will be tidally disrupted and 
accreted at high rate on to the more massive WD.  

We therefore assume in our simulations that a NS will be formed via ECS in the following cases:
\begin{itemize}
\item if at the start of the AGB the He core mass is $1.83~M_\odot~\la~M_{\rm c,BAGB}\la~2.25~M_\odot$  (EIC)
\item if an accreting ONeMg WD reaches $M_{\rm ecs}$  (AIC)
\item if the total mass of two merging or colliding WDs exceeds $M_{\rm ecs}$  (MIC).
\end{itemize} 

For NSs formed via ECS we assume that the accompanying natal kick is
10 times smaller than in the CC case. 
This assumption follows from the results of numerical simulations which find 
that the SASI instability, required by current understanding for the large explosion asymmetry 
in the case of core-collapse supernovae, fails to develop and resulting kicks are significantly smaller
and on overall do not exceed 100 km/s \citep{Buras06,Kitaura06}. 
We will provide the separate statistics
for NSs created by different channels (core-collapse, EIC, AIC or MIC), 
so that if  one of the channels later turns out to have been overestimated 
 it will be easy to recalibrate.

The gravitational mass of the newly formed NS is $1.26\ M_\odot$. This is $\sim 0.9$ of its 
baryonic mass (which cannot exceed the pre-collapse core mass) 
after subtracting the binding energy and is calculated as in \cite{Timmes96_nsbh}:
\begin{equation}
M_{\rm baryon}-M_{\rm grav}=\Delta M = 0.075 M_{\rm grav}^2 \ ,
\end{equation}
where the masses are in solar masses.
As the angular momentum of the binary is conserved, in the case of no kick,
the binary widens. 

The condition for ECS to occur depends essentially on the central density of
the object. As a consequence, a rapidly rotating WD can reach a much higher mass than
the Chandrasekhar limit before the central density becomes high enough for 
electron captures on $^{24}$Mg and $^{20}$Ne to occur. For example, its mass
can be as high as $1.92\ M_\odot$ if the ratio of the WD rotational energy to the
gravitational binding energy is 0.0833 and the star is highly oblate \citep{Dessart06_aic}.
Such rapidly rotating heavy WDs can be formed, e.g., during the coalescence of two WDs 
with total mass exceeding the
Chandrasekhar mass. During such a merger, less than 0.5 per cent of mass will be lost from the system \citep{Guerrero04_sphwd}.
The collapse of a rapidly rotating WD can therefore lead to the formation of a more massive and very 
fast spinning NS.

\subsubsection{Convective and radiative envelopes}

The presence or absence of a deep outer convective envelope
influences a star's behavior during various phases of close
binary interactions (e.g., angular momentum loss via magnetic braking,
tidal interactions, stable or unstable mass transfer).
In particular, it has been pointed out in \cite{Ivanova06_lmxb} that the absence
of a deep outer convective zone in low-metallicity GCs
may explain the preferential formation of LMXBs
in metal-rich clusters \citep{Grindlay93,Bellazzini95,Zepf06}. 

In this study, for main-sequence stars we adopt the metallicity-dependent mass range 
to develop a convective envelope described by \cite[eq.(10) in][]{Bel06_startrack} 
In addition, in our previous work, all giant-like
stars were assumed to have outer convective envelopes. Now 
stars crossing the Hertzsprung gap and stars in the blue loop 
are not assumed to have convective envelopes if their effective temperature
satisfies $\log_{10} T_{\rm eff} > 3.73$.

\subsubsection{Common-envelope phases}

For common-envelope (CE) events we use the standard energy formalism \citep{Livio88_ce}.
In this approach, the outcome of the CE phase depends on the adopted
efficiency for orbital energy transfer into envelope expansion energy $\alpha_{\rm ce}$ and
on the donor envelope central concentration parameter $\lambda$.
The commonly used prescription is $\alpha_{\rm ce}\times\lambda=1$,
and this is adopted in our models as well.
However, for He stars with an outer convective envelope 
we determine $\lambda$ from the set of detailed He star evolution
models calculated by \cite{Ivanova03_he}.
In this case, $\lambda = 0.3 R^{-0.8}$, where $R$ is the radius of the He star in solar radii.
As many He stars end their lives as NSs, the CE evolution of He stars 
is an important channel for the formation of close binaries with NSs.
Indeed, the prescription above was recently used in a study of double neutron star formation
as short $\gamma$-ray burst progenitors \citep{Bel06_grb}.

\subsection{Dynamical events}

\subsubsection{Binary formation via physical collisions with giants}
\label{bf_pc}

We consider in our models the possiblity of (eccentric) binary formation via 
physical collisions involving a red giant (RG).
For binaries formed through NS--RG collisions, the final  binary
separation $a_{\rm f}$ and eccentricity
$e_{\rm f}$ depend on the closest approach distance $p$ \citep{Lombardi06_sph}
and can be estimated using the results of hydrodynamic calculations as

\begin{equation}
e_{\rm f} = 0.88-{\frac {p} {3 R_{RG}}}
\end{equation}
\begin{equation}
a_{\rm f} = {\frac {p} {3.3 (1-e_{\rm f}^2)}}
\label{af_sph}
\end{equation}

\noindent More details on our treatment of these collisions can be found in \cite{Ivanova06_wd}.
In this paper, we refer to this event as to a dynamical common envelope (DCE).

\subsubsection{Collisions and mergers with NSs}

\label{subs_pc}

We assume that the amount of matter accreted by a NS following a merger or collision
depends on the evolutionary stage of the other star.
If a NS collides or merges with a MS or a He MS star, we assume that
it accretes no more than 0.2 $M_\odot$ \citep{Rosswog06_bhms}.
In the case of a collision
with a giant-like star, the maximum accreted mass is 0.01 $M_\odot$ \citep{Lombardi06_sph}.
When the other star is a WD, all the WD mass is accreted.
In essentially all mergers and collisions, the accreted mass must have gone through an  accretion disk,
implying that, if a significant amount of matter is accreted, a recycled pulsar will be formed.

\subsubsection{Triples and their treatment}

Stable hierarchical  triples can be formed through binary--binary encounters. 
Although these triples would be stable in isolation,
it is likely that they will be destroyed during their next dynamical encounter.
As there  are no  developed  population  synthesis  methods  
for handling triple star evolution, we cannot keep these triples in our simulations 
and instead we have to break them immediately after formation into a binary and a single star.
We do this while conserving energy: the energy  required 
to eject the outer  companion is balanced by  shrinking  the 
inner  binary orbit.   The outer  companion is
released  unless the required energy is such that the inner binary  merges.
In the latter  case  the inner  system  is allowed  to  merge  and the  outer
companion is  kept in a new, wider orbit  to form the  final binary
system.

It is possible that in a triple the inner orbit's eccentricity will
be increased via the Kozai mechanism \citep{Kozai62}.
This secular coupling  causes large variations in
the eccentricity  and inclination of  the orbits and  could drive
the inner binary of the triple system to merge before the next dynamical interaction.
The maximum eccentricity in systems with large initial inclinations $i_0$, 
such that $\sin i_0> (2/5)^{1/2}$ ($i_0 \ga 39{\degr}$) is \citep[e.g.,][]{Innanen97_kozai, Eggleton01_triples}:

\begin{equation}
e_{\rm max} \simeq \sqrt{1-5/3 \cos^2(i_0)} \ .
\end{equation}

\noindent The period of the cycle to achieve  $e_{\rm max}$ is  \citep{Innanen97_kozai,Miller02_kozai}

\begin{equation}
\tau_{\rm Koz}  \simeq  \frac{0.42 \ln (1/e_{\rm i}) }{\sqrt{\sin^2(i_0) -0.4}}  
\left ( \frac{m_1+m_2}{m_{\rm o}} \frac{b^3_{\rm o}}{a^3_{\rm i}} \right ) ^{1/2} 
\left ( \frac{b^3_{\rm o}}{ G m_{\rm o}} \right )^{1/2} ,
\label{tkoz}
\end{equation}

\noindent where 
$e_{\rm i}$ is the initial eccentricity of the inner binary,
$m_1$, $m_2$ and $m_o$ are the masses of the inner binary companions and the mass of the outer star,
$a_{\rm i}$ and $a_{\rm o}$ are initial orbital separations for the inner and outer orbits,
and $b_{\rm o}= a_{\rm o}(1 - e_{\rm o})^{3/2}$ is the semiminor axis of the outer orbit.

We  compare the Kozai time-scale  $\tau_{\rm Koz}$ 
with the  collision time $\tau_{\rm coll}$ \citep[computed as in][]{Ivanova05_bf}:

\begin{eqnarray}
\label{tcoll_pd}
\tau_{\rm coll} = 8.5 \times 10^{12} \ {\rm yr} \ \   P_{\rm  td}^{-4/3} M_{\rm tot}^{-2/3} n_5^{-1}
v_{10}^{-1} \times \\ \nonumber
\left( 1+913 {\frac { (M_{\rm tot} + \langle M\rangle )} {k P_{\rm td}^{2/3} M_{\rm  tot}^{1/3} v_{10}^2}}\right) ^{-1}
\end{eqnarray}

\noindent Here $P_{\rm td}$ is the triple period in days,
$M_{\rm tot}$ is the total triple mass in  $M_\odot$,
$\langle M\rangle$ is the mass of an average single star in  $M_\odot$,
$v_{10}=v_{\infty}/(10\,{\rm km/s})$ and $n_5=n/(10^5\,{\rm pc}^{-3})$, where $n$ is the stellar number  density.

If $\tau_{\rm Koz} >  \tau_{\rm coll}$, the Kozai mechanism does not affect the triple
evolution before the next encounter occurs, and we break the triple as describe above.
However,
for triples with $\tau_{\rm Koz} <  \tau_{\rm coll}$, we consider the
circularization time-scale $\tau_{\rm circ}$ \citep[see \S3.3 in][]{Bel06_startrack}. 
The characteristic time-scale for the inner binary separation to decrease through
tidal dissipation is $\tau_{\rm circ}/2 e^2_{\rm i}$ \citep{Mazeh79_triples}.
If  $\tau_{\rm circ}/2 e^2_{\rm i} <  \tau_{\rm coll}$, the inner
binary will shrink until one of the components
overfills its Roche lobe or its separation will small enough to cause the Darwin instability
resulting in a merger \citep{Eggleton01_triples}.
In the opposite case, when $\tau_{\rm circ}/2 e^2_{\rm i} >  \tau_{\rm coll}$,
we assume that mass transfer starts if $e_{\rm max}$ 
is such that at least one of the stars  in the inner binary overfills its Roche lobe at pericentre. 
As a result, triple formation may enhance the formation of mass-transferring 
binaries with a NS.

For sufficiently compact inner binaries, relativistic effects may play an important role.
The relativistic precession period  $P_{\rm pr}$ of the inner binary, 
to first post-Newtonian order, is \citep[see][p.197]{Weinberg72_book}:

\begin{equation}
P_{\rm pr} = \frac{2 \pi c^2 a^{5/2}(1-e_{\rm i}^2)}{3 G^{3/2} (m_1+m_2)^{3/2}}
\end{equation}

\noindent If $\tau_{\rm Koz} > P_{\rm pr}$, Kozai cycles are  strongly 
suppressed \citep{Holman97_planets}.

We expect  that some dynamically formed triples  can also  undergo
significant secular  eccentricity evolution even
for orbital inclinations smaller than the Kozai angle \citep[see
e.g.][]{Ford00_sec}, but here we neglect this possibility.

\section{Radio Pulsar Modeling}
\label{msp-scenario}

\subsection{Slow, non-recycled radio pulsars}

\label{msp-nonrec}

The time over which a pulsar will spin down from its initial period $P_0$ to a current period $P$ is

\begin{equation}
\tau_{\rm MSP} = \frac{P^2-P_0^2}{2} \frac{1}{P \dot P},
\label{eq-msp}
\end{equation}
where $P \dot P=B^2/10^{39}$ [s]. With the assumption that the magnetic field $B$ 
does not decay, $P \dot P$  remains constant with time.

Several observed radio pulsars in GCs are slowly spinning 
($P>0.1$s) and have high $\dot P$s (higher than can be produced by gravitational 
acceleration in the cluster potential) implying relatively high $B$ fields, $>10^{11}$ Gauss.  
These include NGC 6342A \citep{Kerk00_b1718}, NGC 6624B \citep{Biggs94}, 
and NGC 6440A \citep{Lyne96}.  Their inferred characteristic ages range from 
$10^7$ yr to $2\times10^8$ yr, compared to $10^{11} - 10^{13}$ yr for ``normal'' MSPs. 
These pulsars cannot have been formed in their current state through core-collapse SNe, which only occur within the first ten million years.
 The inferred birthrate for this class of objects is similar to that of ``normal'' MSPs, 
but observationally they seem to be concentrated in the densest clusters, and 2 of 3 
are isolated pulsars.  

A strong magnetic field of $>10^{11}$-gauss can be produced by the collapse, 
assuming flux conservation, of a typical magnetic WD with $B\sim 10^6$-gauss.
Observed magnetic WDs typically have higher masses than nonmagnetic WDs\citep{Wick05}, 
indicating that they may dominate the population of WDs that undergo AIC. 
 Stellar dynamo theory supports the link between strong magnetic fields and massive WDs \citep{ThompsonDuncan_mf}.
It is also possible that a  strong magnetic field may be generated during the merger of a WD pair,
where one or more WDs had a strong magnetic field, leading to the production 
of $10^{12}$-gauss magnetic fields, or even magnetar $B$ field strengths of  
$10^{14}\div10^{15}$-gauss \citep{King01_mic, Levan06_magnetars, Chapman06_magnetars}.
MSPs formed with such high magnetic fields will have relatively short lifetimes, 
and thus will make up only a small portion of the observed globular cluster MSP population.
In our simulations, we do not assign a specific magnetic field to formed neutron stars.
However we keep a record of their formation and therefore can test whether the population of
AIC and MIC NSs is formed at the required rates to explain the
population of slow GC radio pulsars,
assuming that their life-time is only $10^8$ years.

We refer to young, recently formed  pulsars that have not yet been recycled as ``1a'' and ``1b'' in  Fig.~\ref{fig-msp-scenario}.
As a standard picture, we consider that the origin of a NS does not affect its further evolution
if a MT event occurred in a dynamically formed binary (c.f. post-AIC evolution), but
we note that it is possible that MIC NSs might have a higher than usual magnetic field
and their subsequent evolution might be different -- they can be seen only for a shorter time after their formation
as young slow pulsars and they might be not able to accrete and be spun up.

\subsection{{Mild recycling in post-AIC systems}}

\label{msp-mild-rec}

Very little is known about whether a magnetic field is always reduced during mass
accretion, independently of its initial strength. 
E.g., the UCXB and 7.7 second X-ray pulsar 4U 1626-67 has a strong ($4\times10^{12}$-gauss) magnetic field \citep{Orlandini98_4u_mf}.  
On the other hand, known magnetic fields  in accreting MSPs are usually $10^7$- $10^9$ gauss
(e.g., $1-5\times10^8$-gauss in XTE J1751-305, \citet{Salvo03}; $<3\times10^8$-gauss in IGR J00291+5934 \citet{Torres07}; 
$\la10^9$-gauss in XTE J0929-314, \citet{Galloway02}) and are consistent 
with the magnetic fields of MSPs.
Even though 4U 1626-67 is evolving through the accretion phase and has passed the stage when the
MT rate is maximum, the NS has kept a high magnetic field and has not been spun up to millisecond periods. 
To explain this, it has been suggested that the NS in 4U 1626-67 was recently formed via AIC \citep{Yungelson02}. 
Several radio pulsars in the Galaxy possess relatively high $B$ fields 
and long periods, with low-mass companions in low-eccentricity orbits \citep{Breton07}.  
They argue that these systems can be well-explained through AIC.

It is unclear why post-AIC binaries may yet fail to produce MSPs.  
Reasons could include the high initial $B$ field and moderately fast spin of newly-formed AIC NSs, 
which could disrupt spin-up; or simply that after sufficient accretion 
onto the WD to reach the Chandrasekhar limit, there is usually not enough remaining mass 
in the companion to spin the NS up to millisecond periods.   
We will show below that producing MSPs in the same binaries in which AIC 
occurred contradicts observations.  For our final results, we therefore adopt 
that NSs formed via AIC or MIC have short lifetimes ($10^8$ years), 
and that NSs recycled in their original AIC systems are only mildly recycled, also with lifetimes 
of $10^8$ years (case ``2a'' in ~\ref{fig-msp-scenario}).

\subsection{{Recycling via stable MT}}

\label{msp-rec-mt}

A common understanding of recycled pulsar formation is that 
the NS is recycled through disk accretion  and has its magnetic fields reduced 
during the accretion phase \citep{vandenH84_msp}. 
Since the characteristic lifetime of a MSP is $10^{11}-10^{13}$ years and is significantly larger than
a GC life-time ($10^{10}$ years), we do not consider in our simulations the evolution of MSPs 
towards a death zone after they are formed.

The amount of material that needs to be accreted on to a NS to form a MSP is however not 
 firmly established.  In our simulations, we record for further analysis any event through which
a NS acquires at least 0.01~$M_\odot$. 
We also distinguish mass gain through steady accretion from that  
which can result from a physical collision with a red giant 
(DCE); a CE event during binary evolution; or a merger or physical collision with a MS star or a WD.

The connection between stable mass-transfer and the binary 
MSP production in the field seems to be reasonably well-understood and is reviewed by \cite{Deloye07_lmxb}. 
We outline only the main branches:

\begin{itemize}
\item{If the donor is a RG or an asymptotic-giant branch companion, then the resulting binary MSP
has a wide orbital separation (``4b'' at the Fig.~\ref{fig-msp-scenario}).}

\item{If the MT occurred in a NS-MS binary, then the evolution depends on the period at the start of the MT.
Some systems will evolve towards shorter periods(``3a''  at the Fig.~\ref{fig-msp-scenario}), 
and others to longer (``4d''  at the Fig.~\ref{fig-msp-scenario})
\citep[for a review of the current understanding of NS-MS LMXBs evolution see][]{Deloye07_lmxb}.
The progenitors to the short-period binary MSPs evolve slowly enough that they may be 
detected as (bright or quiescent) LMXBs; systems with these orbital periods are indeed seen in globulars \citep{Heinke05b,Altamirano07}.
}

\item{
If a binary system with a NS has evolved through a CE resulting in a significant shrinkage
of a binary orbit, MT with a WD donor can start.
This leads to the formation of an ultra-compact X-ray binary (UCXB) (``3b''  at the Fig.~\ref{fig-msp-scenario}).
}
\end{itemize}

We do not discuss here the case of thermal-time scale MT that occurs with an intermediate-mass donor
($2\le M_{\rm d}\le 4M_\odot$) in a close orbit at the start of the MT (0.3-2 days), as we find it unlikely 
for a NS in a GC environment to acquire an intermediate mass donor \citep[c.f.][]{Rasio00}.
We will return to this point later in \S8 and only note here that the evolutionary life-time of 
an intermediate mass donor is not much longer than the cluster age when ECS NSs are formed.

Even if the MT leads to spin-up,
a NS does not appear as a radio MSP immediately even if has accumulated
sufficient mass.  While the MT rate remains high, systems continue to appear as LMXBs. 
In general, LMXBs with a likely MS donor do not show up as radio MSPs
while the companion mass is above $\sim 0.2 M_\odot$ \citep[see Fig.~1 in][]{Deloye07_lmxb}.

\subsection {{Short-period radio bMSPs}}

\label{mt-bmsp}
 
There are discrepancies between the observed orbital periods of 
short-period radio binary MSPs (bMSPs) with very low-mass companions and the modeled periods.
Using a standard mass-radius relation for WDs during the MT evolution of a NS-WD binary gives
periods several times smaller than those of observed bMSPs with companions of similar masses. 
It has been shown that even the consideration of more realistic WD models does not change the result strongly --
hotter donors don't, by themselves, produce the long periods needed to explain the short-period bMSPs \citep{Deloye03_wd,Deloye07_lmxb}.
If the mass-losing companion is a WD, some poorly understood process must be invoked 
-- e.g., strong tidal heating of the WD during MT \citep{Rasio00}. 
In that case, it is not possible to accurately predict the bMSP periods during the MT or at its end.

There is no clear termination point for MT from a degenerate donor. 
As a result, the MT continues uninterrupted and within the Hubble time 
the companion mass can reach below $0.01\ M_\odot$.
No observed radio bMSPs have been shown to have such low-mass companions, in 
globular clusters or the field.

Several  accretion-powered MSPs  with $P_{orb}<60$  minutes  have been
identified,  consistent  with (partly)  degenerate  He  or C/O  donors
\citep[e.g. ][]{Krimm07},  and several are known  in globular clusters
\citep[e.g.][]{Dieball05}.  However, no radio bMSPs have been observed
with such short  periods.  These systems will spend  Gyrs as transient
LMXBs,  with  continually  decreasing  mass transfer  rates  
(``3b'' and ``6a''  in Fig.~\ref{fig-msp-scenario}).  
It is  possible that  the UCXBs  in clusters  remain  in this
configuration, never  turning on as millisecond pulsars,  for a Hubble
time.  A sizable number of such systems could remain hidden in globular
clusters if  their accretion outburst intervals are sufficiently  long.
Alternatively, partly degenerate (due to irradiation) donors will have
a  minimum companion  mass for  a given  temperature,  indicating that
evaporation of  the donor star and production  of isolated millisecond
pulsars is a likely outcome \citep{Bildsten02,Deloye03_wd}.   Other
possibilities include very strong tidal heating  of the  companion
leading to strong orbital  expansion, producing the low-mass eclipsing
pulsars \citep{Rasio00}, or spin-down of the NS due to decreasing mass
transfer rates \citep[see discussion in ][]{Deloye07_lmxb}.  
Or, current pulsar detection algorithms may not yet be sufficiently sensitive to detect such short-period systems \citep[e.g. ][]{Hessels07}.   
We do not know
which  of these  fates befalls  ultracompact LMXBs,  although  we know
observationally that ultracompact radio bMSPs (``6a'' in Fig. 1) have not
yet been seen.   We therefore count  the numbers  of ultracompact  systems that
have evolved past 55 minutes  \citep[the longest orbital period yet observed
for  an ultracompact LMXB,][]{Krimm07}  and tabulate  them as
ultracompact  MSPs, but  do not  plot their  final binary  periods and
masses; their true final fate is uncertain.

In the case of  MT in a NS-MS binary, if the resulting binary has a 
short period, it will have a (partly) degenerate hydrogen companion. 
Such a companion is larger than a hydrogen-poor WD, and better matches the observed periods 
of known low-mass eclipsing radio bMSPs than systems with WD donors.  
It is thought that the MT becomes unstable and leads to a merger of the two stars 
when the hydrogen-rich companion becomes less massive than  $\sim0.016 M_\odot$ \citep[for analytic estimates see][]{King05_bw}, 
producing a single MSP (``4a'' in fig.~\ref{fig-msp-scenario}).  
However, current binary evolution codes lack the equation of state appropriate for low temperatures.  
Therefore, as of today, there are no direct mass transfer calculations that can proceed 
below donor masses of 0.04 Msun \citep{vanderSluys05_mb1,vanderSluys05_mb2}, 
which is well above the masses of observed bMSPs with very low-mass companions.
The main reason is that an equation of state that is appropriate for
extremely low-mass stars \citep[e.g., ][]{1995ApJS...99..713S} is not yet incorporated
in current binary evolutionary codes.

Our code evolves a NS-MS binary down to very low masses assuming a simple 
mass-radius dependence for a hydrogen-rich degenerate donor.
We find that this evolution ends as a merger when the donor mass is $\sim0.02 M_\odot$,
where this value varies a bit depending on the initial donor mass.
While evolving, the system appears in a globular cluster as first a qLMXb and then as a bMSP for substantial lengths of time.
The values of the donor mass when the instability occurs, and the orbital periods during the qLMXB 
stage, could use refinement when improved evolutionary simulations 
are available.

\subsection{{Recycling during fast (``dynamical'') events}}

\label{msp-fast-rec}

\begin{figure*}
\includegraphics[height=.9\textheight]{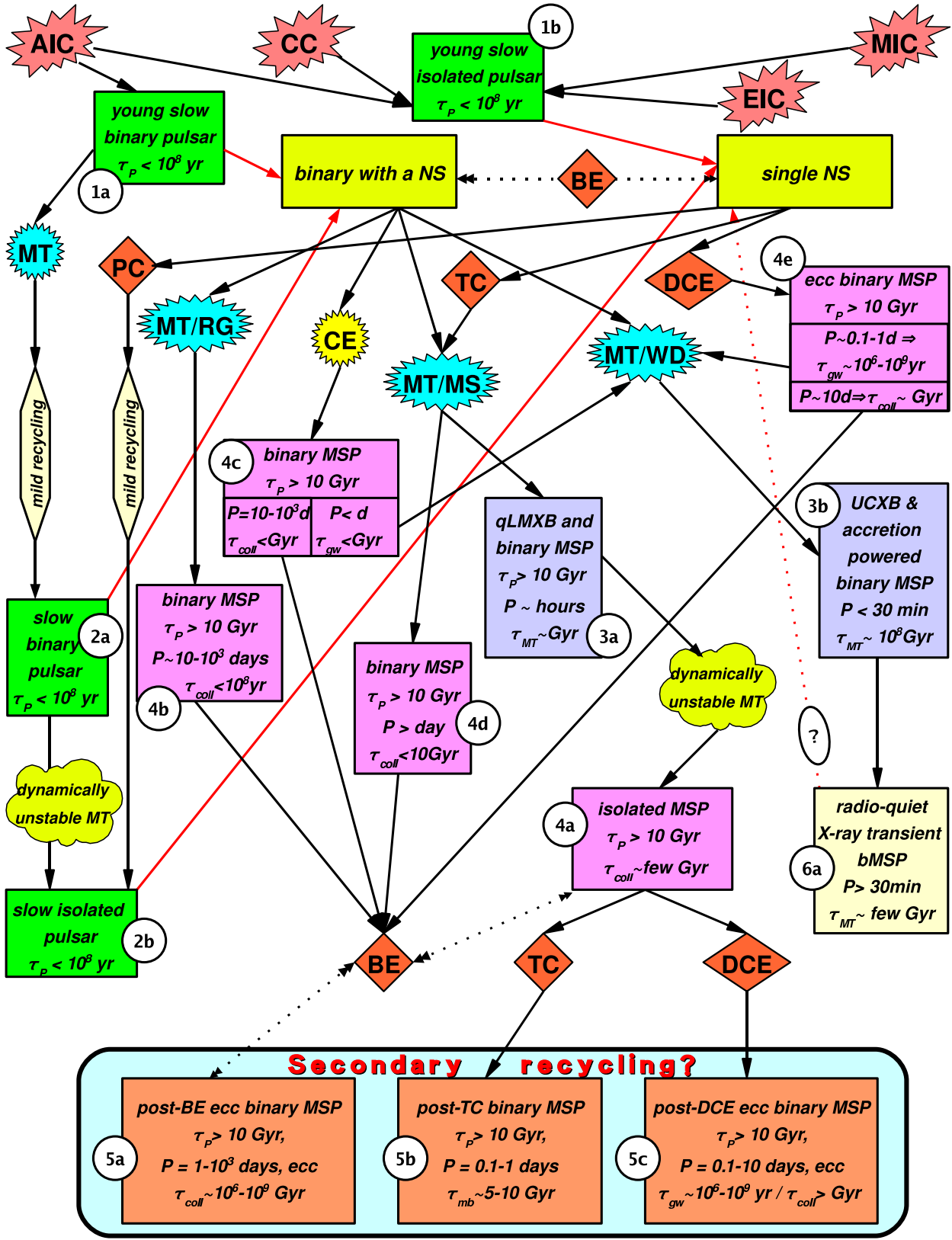}
\caption{
The scenario for the formation and evolution of MSPs in  globular clusters.
$\tau_{\rm P}$ is the life-time of a radio pulsar after formation.
$\tau_{\rm MT}, \tau_{\rm mb}, \tau_{\rm gw}, \tau_{\rm coll}$ indicate which time-scale determines the lifetime of the indicated binary:
$\tau_{\rm MT}$ -- while MT lasts, $\tau_{\rm mb}$ and $\tau_{\rm gw}$ -- until magnetic braking or 
gravitational radiation merges the binary, $\tau_{\rm coll}$ -- before the next dynamical encounter.
The final fate of radio-quiet X-ray transient bMSPs (``6a'') is not determined.
\label{fig-msp-scenario}
}
\end{figure*}

We consider here mass gain which can result from a physical collision with a red giant 
(DCE,case ``4e'', see Fig.~\ref{fig-msp-scenario}) ; a CE event during binary evolution (case ``4c''); 
or a merger or physical collision with a MS star or WD (``2b'' at the Fig.~\ref{fig-msp-scenario}).

During a physical collision with a giant, the NS retains  
$\le 0.01 M_\odot$ of the giant's envelope \citep{Lombardi06_sph}.
This matter has high specific angular momentum and the resulting accretion  must proceed through an  accretion disk,
implying that a recycled pulsar can be formed.

A CE event is also likely to lead to spin-up. As an example, it is believed that 
the double neutron star system PSR~J0737-3039 is a post-common envelope remnant \citep{Dewi04},
where the first-formed NS has been spun up during a CE event to millisecond periods.  

It is less clear whether possible accretion as a result of a merger leads to recycling. 
Such mergers have been considered to produce only  mild spin up \citep{Lyne96}.
For a standard scenario we assume that physical collision of NSs with other stars 
leads only to the formation of short-living slow (isolated) pulsars.

\subsection{{Post-recycling phase}}

\label{msp-postrec}

Once recycled, a single or binary MSP can participate again in dynamical encounters, and form new binary systems
whose properties will depend on the type of binary formation: binary exchange (``5a''),
tidal capture (``5b'') or another DCE (``5c''). This type of binary millisecond pulsars, along with
post-DCE ``4e'', should not be formed in the field population. 
Newly dynamically formed binaries with MSPs may evolve again through stable MT.

\section{Globular cluster models}
{

\begin{table*}
\caption{Models.}
\label{tab-model}
\begin{tabular}{@{}l c c c c c c c c c c c}
\hline
 & $n_{\rm c}$ & $t_{\rm rh}$ & $Z$ & $\sigma_1$  & $v_{\rm esc}$ & $f_{\rm b}^0$  & $\alpha_{\rm CE}\lambda$ & MB & ${ M_{\rm cluster}}$ & Runs (number of stars)\\
\hline
standard   & $10^5$ & 1.0  & 0.005 & 10 & 40 & 100\%  & 1.0 & IT03 & $2\times 10^5 M_\odot$ & { 5 with $10^6$}, { 5 with $2\times10^6$}\\
metal-poor & $\cdot$ &  $\cdot$ &  0.0005 &  $\cdot$ &$\cdot$ &  $\cdot$    &  $\cdot$ &  $\cdot$   &  $\cdot$   &  { 5 with $10^6$} \\
high-dens   &  $10^6$ &  $\cdot$ &  $\cdot$ &  $\cdot$ &$\cdot$ & $\cdot$    &  $\cdot$ &  $\cdot$  &  $\cdot$  & { 5 with $10^6$} \\
med-dens   &  $10^4$ &  $\cdot$ &  $\cdot$ &  $\cdot$ &$\cdot$ & $\cdot$    &  $\cdot$ &  $\cdot$  &  $\cdot$  & { 5 with $10^6$} \\
low-dens   &  $10^3$ &  $\cdot$ &  $\cdot$ &  $\cdot$ &$\cdot$ &  $\cdot$   &  $\cdot$ &  $\cdot$   &  $\cdot$ & 5 with $10^6$ \\
low-$\sigma$   &  $\cdot$ &  $\cdot$ & $\cdot$ & 5 & 20 &  $\cdot$   &  $\cdot$ &  $\cdot$   &  $\cdot$ & { 5 with $10^6$} \\
long-$t_{\rm rh}$ &  $\cdot$  &  3.0  & $\cdot$ & $\cdot$ & $\cdot$ &  $\cdot$   &  $\cdot$ &  $\cdot$   &  $\cdot$ &  5 with $10^6$ \\
BF05      &  $\cdot$ &  $\cdot$ &  $\cdot$ &  $\cdot$ &  $\cdot$ & 50\%  &     $\cdot$ &  $\cdot$  &  $\cdot$   & 5 with $10^6$  \\
fast-MB    & $\cdot$ &  $\cdot$ &  $\cdot$ &  $\cdot$ &$\cdot$ &  $\cdot$   &  $\cdot$ &  RVJ  &  $\cdot$   & 5 with $10^6$ \\
CE-reduced   &  $\cdot$ &  $\cdot$ &  $\cdot$ &  $\cdot$ & $\cdot$ &  $\cdot$  &  0.1 &  $\cdot$ &  $\cdot$   & 5 with $10^6$ \\
oldkicks   &  $\cdot$ &  $\cdot$ &  $\cdot$ &  $\cdot$ & $\cdot$ &  $\cdot$  & $\cdot$  &  $\cdot$ &  $\cdot$   & { 5 with $10^6$} \\
47 Tuc    &  $2.5\cdot10^5$ &  3.0 &  0.0035 &  11.5 & 57 &$\cdot$ &  $\cdot$ &  $\cdot$  &  $ 10^6 M_\odot$  & { 5 with $2\times10^6$}\\
Terzan 5 & $8\cdot 10^5$ &  0.93 &  0.013 &  11.6 &  49 &   $\cdot$ & $\cdot$ &  $\cdot$   &  $3.7\times 10^5 M_\odot$ & { 5 with $1.7\times10^6$} \\
\hline
\end{tabular}

\medskip
Notations: $n_{\rm c}$ -- core number density ($pc^{-3}$); $t_{\rm rh}$ -- half-mass relaxation time (Gyr); $Z$ -- metallicity;
$\sigma_1$ -- one-dimensional velocity dispersion (km/s); $v_{\rm esc}$ -- central escape velocity (km/s);
 $f_{\rm b}^0$ -- initial binary fraction (see text for corresponding binary fraction of hard binaries);
$\alpha_{\rm CE}\lambda$ -- CE efficiency; MB  -- adopted prescription for  magnetic braking (IT03 is from \cite{Ivanova03_mb} and RVJ is from
\citet{RVJ}); ${ M_{\rm cluster}}$ -- all the results for this model are scaled to this adopted cluster mass.
``Oldkicks'' model uses the distribution of natal kick velocities from \citep{Arzoumanian02ApJ_kicks}.
 Symbol ``$\cdot$'' means that the value is the same as in the standard model.
\end{table*}

Our ``standard'' cluster model  has an initial binary fraction of 100 per cent.
The distribution of initial binary periods is constant
in the logarithm between contact and $10^7$~d and the eccentricities are distributed thermally.
We emphasize that about 2/3 of these binaries are soft initially (the initial
binary fraction for hard binaries is about $20$ per cent if the 1-D velocity dispersion is 10 km/s)
and most very tight binaries are destroyed through evolutionary mergers.
Our initial binary fraction is therefore comparable to the initial binary
fractions that are usually adopted in $N$-body simulations, 10-15 per cent in hard binaries
(soft binaries are short-lived in dense clusters but slow down the calculations considerably).
For a more detailed discussion of the choice of primordial binary fraction, see \cite{Ivanova05_bf}.

For single stars and primaries we adopt the broken power law initial mass function (IMF) of \citet{Kroupa02}
and we assume a flat mass-ratio distribution for secondaries. The initial
core mass is 5 per cent of the total cluster mass and no initial mass segregation is assumed 
(an average object in the core is as massive as an average cluster star).
At the age of 11 Gyr, which we adopt as a typical age of a Galactic GC, 
the mass of such a cluster in our simulations is $\sim 2.2\times 10^5\,M_\odot$
for the metallicity of a typical metal-rich cluster ($Z=0.005$) and is comparable to the mass of typical GCs 
in our Galaxy.
In the case of a metal-poor cluster (Z=0.0005), the mass is  $\sim 2.5\times 10^5\,M_\odot$.
To make a comparison between cluster models of different metallicities
we will show all results scaled to a cluster mass $2\times 10^5\,M_\odot$.
As the metal-rich case observationally is more important -- most bright LMXBs and MSPs detected so far are located there -- 
we take for our standard model the value $Z=0.005$.

For our standard model we fix the core density at $n_{\rm c}=10^5 \ {\rm pc}^{-3}$.
The link to an observed luminosity density can be found using the number density-to-mass density ratio, which in  
our simulations is $\sim 2 \ M_\odot^{-1}$ at the ages of 7-14 Gyr, and mass-to-light ratio, 
which on average is $\sim 2.3\ M_\odot/L_\odot $ and is $2\ M_\odot/L_\odot$ for 47~Tuc \citep{PryorMeylan93}. 
We adopt a half-mass relaxation time  $t_{\rm rh}=1$ Gyr \citep{Harris96}.
The characteristic velocities are computed for a King model with dimensionless
central potential $W_0=7$ and the total model mass, giving
a one-dimensional velocity dispersion
$\sigma_1=10$ km/s and a central
escape velocity  $v_{\rm esc}=40$ km/s.  
If, after an interaction or SN explosion, an object in the core acquires a velocity
higher than the {\em recoil velocity\/} $v_{\rm rec}=30$ km/s, the object is moved from the core to the halo.
The ejection velocity for objects in the halo is $v_{\rm ej,h}=28$ km/s.
This standard model represents a typical dense globular cluster in  our Galaxy.

In  addition to this ``standard'' model
we also considered cluster models with the following modifications:
\begin{itemize}
\item a metal-poor cluster with $Z=0.0005$ -- ``metal-poor'';
\item different central densities: $n_{\rm c}=10^6 \ {\rm pc}^{-3}$ , $n_{\rm c}=10^4 \ {\rm pc}^{-3}$  
and  $n_{\rm c}=10^3 \ {\rm pc}^{-3}$ --  ``high-dens'', ``med-dens'' and ``low-dens'' respectively; 
\item a cluster with smaller velocity dispersion and smaller ejection velocity -- ``low-$\sigma$'';
\item shorter half-mass relaxation time ``long-$t_{\rm rh}$'';
\item reduced initial binary fraction, $f_{\rm b}^0=50$ per cent  -- ``BF05'';
\item magnetic braking prescription of \cite{RVJ} -- ``fast MB'';
\item reduced efficiency of the common envelope, $\alpha_{\rm CE}\lambda=0.1$ -- ``CE-reduced'';
\item natal kick distribution is doubled peaked Maxwellian as in \cite{Arzoumanian02ApJ_kicks} and
NSs are produced only via core-collapse and double WD collisions (MIC), but not through EIC, AIC 
or evolutionary issued MIC -- ``oldkicks''
\item 47~Tuc-type cluster, characterized by  a higher density
$n_{\rm c}=10^{5.4}\,{\rm pc}^{-3}$ (with M/L=2 as in \citep{PryorMeylan93} it corresponds to 
$\rho_{\rm c}=10^{4.8}\,L_\odot \ {\rm pc}^{-3}$ as in \cite{Harris96}) , metallicity $Z=0.0035$,
$t_{\rm rh}=3$ Gyr \citep{Harris96}
$\sigma_1=11.5$ km/s \citep{PryorMeylan93}, $v_{\rm esc}=57$ \citep{Webbink85_gc}, with the recoil velocity
of 51 km/s and $v_{\rm ej,h}=32$ km/s  (King model $W_0=12)$) -- ``47 Tuc'';
\item Terzan 5 cluster, $n_{\rm c}=10^{5.9}\,{\rm pc}^{-3}$ ($\rho_{\rm c}=10^{5.23}\,L_\odot \ {\rm pc}^{-3}$ 
as in \cite{Heinke03a}), metallicity $Z=0.013$ \citep{OrigliaRich04_ter}, 
$t_{\rm rh}=0.9$ Gyr \citep{Harris96} 
$\sigma_1=11.6$ km/s, $v_{\rm esc}=49.4$ \citep{Webbink85_gc}, 
with the recoil velocity of 39 km/s and $v_{\rm ej,h}=29$ km/s  (King model $W_0=8.3$)   -- 
``Terzan 5''.\footnote{The recent study of \citet{Ortolani07} suggests a shorter distance, and thus higher central density for Terzan 5 
($\rho_{\rm c}=10^{5.61} L_\odot$/pc$^3$).  
If this study is correct, then Terzan 5's characteristics may be more closely approximated by our ``high-density'' cluster.}
\end{itemize}
For the complete list of models see Table~\ref{tab-model}.

We expect that only a few LMXBs  can be formed in every cluster with
a typical mass  of 200,000 $M_\odot$ through 11  Gyr of dynamical evolution,
and  therefore we  perform 5  runs  for each  model with  $10^6$
initial  stars  and   for  the standard  models  -   five  runs  with
$2\times10^6$ stars,  to check that  statistics for rare  objects like
LMXBs are  not different  when the resolution is increased 
(for most of our results
presented below we did not find statistical differences between high and low resolution
models, and fluctuations for the rate events are smaller in the case of high resolution). 
For Terzan~5 and 47~Tuc models, though, we do all runs at
high   resolution.   In   particular,   in  the   case  of   Terzan~5
$1.7\times10^6$ stars  provide at  11 Gyr the  mass comparable  to the
estimated observed mass  of 370,000 $M_\odot$.  In the  case of 47 Tuc
the reason for  a higher resolution is that the  relative core mass is
smaller than in the case of  a standard cluster, due to longer $t_{\rm
rh}$.

}

\section{Neutron star formation and retention}

\subsection{Formation}

\begin{table}
\caption{Production of NSs, field population.}
\label{tab-ns-nod}
\begin{tabular}{@{}l  c c c c c c c}
\hline
 & CC & EIC & AIC & MIC  & $N^{\rm cc}_{40}$ & $N^{\rm tot}_{40}$ & $N_{\rm NS}$ \\
\hline
single \\
 Z=0.0005& 3354 & 594& - & - & 2.1 & 251& 584848\\
 Z=0.001  & 3255 & 581 & - & - & 1.9 & 245 & 565168\\
 Z=0.005 & 2833 & 570 & - & - & 1.2 & 243 & 498513\\
 Z=0.02 & 2666 & 400 & - & - & 1.2 & 172 & 424696\\
oldkicks & 3104 & - & - & - & 19.3 & 21.9 & 339669 \\
binary \\
 Z=0.0005  & 3079 & 545 & 59 & 14  & 19 & 329  & 249775\\
 Z=0.001  & 3056 & 553 & 60 & 15  & 15 & 332  & 242601\\
 Z=0.005  & 2750 & 576 & 58 & 16  & 14 & 335  & 219236\\
 Z=0.02  & 2463 & 406 & 33 & 20  & 12 & 240  & 180582\\
 oldkicks & 2929 & - & - & - & 58 & 58 & 237705 \\
\hline
\end{tabular}
\medskip
Notations for channels -- CC - core-collapse supernova; EIC - evolution induced collapse via electron-capture supernova;
AIC - accretion induced collapse; MIC - merger induced collapse.
Numbers are scaled per 200,000 $M_\odot$ stellar population mass at the age of 11 Gyr.
$N^{\rm cc}_{40}$ and $N^{\rm tot}_{40}$ are the retained numbers of 
NSs formed via core-collapse and through all channels, if the escape velocity is 40 km/s. 
$N_{\rm NS}$ are the total numbers of formed NSs in simulated populations.
\end{table}

To estimate how many NSs can be formed and retained in GCs,
we first studied stellar populations without dynamics,
considering separately populations with only primordial single stars and only primordial binaries.
We considered several metallicities -- from solar (as in the Galactic field) to that of a typical 
metal-poor  cluster: Z=0.02, 0.005, 0.001, 0.0005. 
In addition, for comparison with studies done by different researchers in the past, we examined  a model (with Z=0.005) with the previously 
widely accepted natal kick distribution, 2 Maxwellians with a lower peak for kick velocities at 90 km/s \citep{Arzoumanian02ApJ_kicks}.
In this model we adopt that NSs are produced only via core-collapse or through collisional WD mergers, no evolutionary ECS.
The results for NS production via different channels are shown in Table~\ref{tab-ns-nod}.

The production of NSs via core-collapse SNe (CC NSs), per unit of total stellar mass 
at 11 Gyr, decreases as metallicity increases. 
The difference between the Galactic field case and a metal-poor case is about 20 per cent, while the difference 
between metal-rich and metal-poor clusters (Z=0.005 and Z=0.0005) is about 10-15 per cent,
for primordial single or binary populations.

EIC in single stars for metal-rich populations occurs in stars of higher masses.
The range of initial masses is, however, the same for both populations and is about $0.6 M_\odot$.
As a result, the number of NSs produced via EIC (EIC NSs) from the population of single stars is smaller
in the case of a Galactic field population than in the case of a GC population by 30 per cent,
while the difference between metal-rich and metal-poor clusters is only 5 per cent. 
This is in complete agreement with the adopted power law for the IMF.
In binaries, the mass interval where EIC could occur is expanded compared to single stars
of the same metallicity.

NSs produced via AIC and MIC also 
have similar production rates in metal-rich and metal-poor populations.
We note that MIC gives the smallest contribution to the production of NSs in populations evolved without dynamics.

It can be seen that once we adopt that a NS can be created via an ECS, the number of formed NSs via CC is
depleted -- some stars would be evolving via core-collapse, but have experienced the condition for a EIC
before that.

\begin{figure}
\includegraphics[height=.35\textheight]{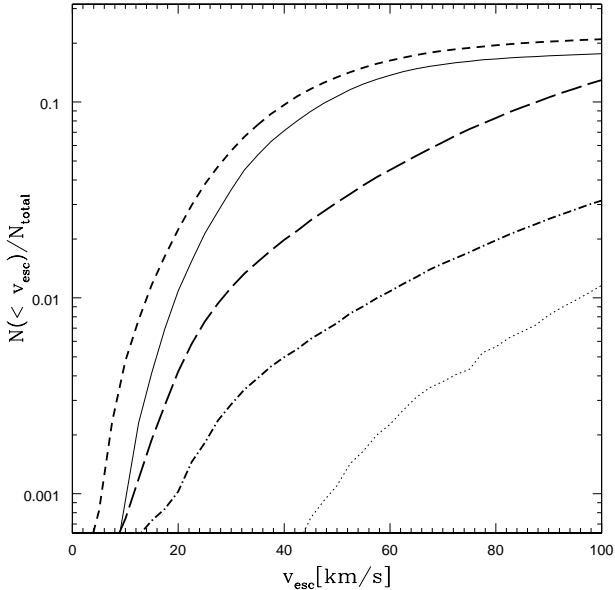}
\caption{
The retention fractions as a function of escape velocity (for stellar evolution unaffected by dynamics). 
Dotted and dash-dotted lines show the retention fractions for single and binary populations,
core-collapse NSs only. Solid and short-dashed lines show the total retention fractions for single and binary populations,
all NSs. Long-dashed line shows the total retention fraction of a binary population with the old kick distribution.
\label{fig-ns-ret}
}
\end{figure}

\subsection{Retention}

In Fig.~\ref{fig-ns-ret} we show the retention fraction of NSs at formation 
as a function of fixed escape velocity the whole population, considering populations of metal-poor
primordial singles and binaries; CC NSs and the whole NS population
are shown separately. 
It can be seen that, with the adopted natal kick distribution, 
CC NSs from a single star population are rarely retained (see also Table~\ref{tab-ns-nod}).
In a population of primordial binaries, the fraction of retained CC NSs is higher than
for single stars. None the less, the resulting number of
retained CC NSs is significantly smaller than we get from the other NS formation channels.
We conclude that if indeed EIC, AIC or MIC are not accompanied by a large natal kick, 
the total number of retained NSs is high and only a negligible number of CC NSs is present.
As the EIC/AIC NS masses are only $1.26 M_\odot$, this may lead to a NS mass function
skewed towards low values.

With the old distribution for natal kick velocities, reasonable numbers of NSs can be produced and
retained in a typical GC, while a massive cluster like 47~Tuc could retain as many as 600 NSs 
(see also Fig.~\ref{fig-ns-ret}). This result agrees with the retention fractions obtained in \citet{Pfahl02}.

\begin{table*}
\caption{Retained NSs and their locations at 11 Gyr, for different globular cluster models.}
\label{tab-ns-dyn}
\begin{tabular}{@{}l c c c c c c c c}
\hline
model & \multicolumn{4}{c}{Core} & \multicolumn{4}{c}{Halo} \\
& CC & EIC & AIC & MIC & CC & EIC & AIC & MIC  \\
\hline
standard & $ 3.5\pm1.5 $& $65.6\pm9.3 $& $21.1\pm3.7 $& $10.3\pm4.0 $& $ 4.8\pm1.7 $& $91.4\pm 11 $& $18.0\pm3.9 $& $ 9.7\pm1.7 $\\ 
metal-poor & $ 3.6\pm1.7 $& $71.5\pm 12 $& $30.4\pm3.2 $& $ 8.4\pm4.5 $& $ 2.3\pm1.1 $& $91.7\pm9.2 $& $20.9\pm4.8 $& $10.9\pm3.8 $\\ 
high-den & $ 3.3\pm1.7 $& $65.5\pm7.8 $& $42.7\pm5.8 $& $20.7\pm6.1 $& $ 4.2\pm3.3 $& $ 111\pm 12 $& $29.0\pm5.2 $& $12.8\pm3.4 $\\ 
med-den & $ 3.3\pm2.1 $& $63.7\pm5.2 $& $18.6\pm4.5 $& $ 6.4\pm3.0 $& $ 4.8\pm1.5 $& $85.6\pm8.0 $& $14.3\pm3.1 $& $ 9.4\pm3.7 $\\ 
low-den & $ 3.2\pm1.3 $& $65.1\pm5.9 $& $14.2\pm3.3 $& $ 9.8\pm0.9 $& $ 3.8\pm1.3 $& $91.2\pm 11 $& $15.4\pm2.1 $& $ 7.0\pm2.0 $\\ 
lowsig & $ 0.5\pm^{0.8}_{0.5} $& $15.5\pm3.2 $& $ 5.5\pm2.3 $& $10.7\pm2.5 $& $ 0.9\pm^{1.1}_{0.9} $& $19.4\pm4.2 $& $ 4.3\pm1.7 $& $11.5\pm1.8 $\\ 
longtrh & $ 1.4\pm1.0 $& $30.0\pm4.7 $& $ 9.3\pm2.1 $& $ 3.0\pm1.2 $& $ 5.3\pm0.9 $& $ 124\pm 15 $& $22.5\pm5.5 $& $15.5\pm3.3 $\\ 
BF05 & $ 1.8\pm1.2 $& $64.1\pm5.0 $& $15.0\pm3.1 $& $ 6.7\pm2.5 $& $ 3.9\pm1.0 $& $88.6\pm6.7 $& $11.4\pm2.7 $& $ 6.2\pm1.7 $\\ 
fast-MB & $ 3.8\pm1.6 $& $77.3\pm6.0 $& $22.9\pm3.3 $& $10.6\pm4.3 $& $ 5.1\pm1.9 $& $93.7\pm6.9 $& $17.2\pm3.3 $& $ 8.3\pm3.2 $\\ 
CE-reduced & $ 2.9\pm1.9 $& $72.2\pm5.6 $& $12.7\pm2.8 $& $ 9.4\pm1.9 $& $ 3.6\pm2.6 $& $99.2\pm4.0 $& $ 4.5\pm1.7 $& $11.7\pm1.9 $\\ 
oldkicks & $14.6\pm3.9 $& $ 0.0\pm0.0 $& $ 0.0\pm0.0 $& $ 0.3\pm^{0.4}_{0.3} $& $21.3\pm4.0 $& $ 0.0\pm0.0 $& $ 0.0\pm0.0 $& $ 0.0\pm0.0 $\\ 
Tuc-47 & $11.8\pm7.5 $& $ 215\pm 20 $& $91.8\pm 23 $& $27.7\pm8.0 $& $38.3\pm7.9 $& $ 598\pm 41 $& $ 102\pm4.9 $& $69.3\pm 12 $\\ 
Terzan-5 & $ 5.1\pm1.4 $& $ 132\pm8.9 $& $91.9\pm 11 $& $47.8\pm4.6 $& $ 8.9\pm1.4 $& $ 143\pm 10 $& $38.0\pm6.6 $& $22.7\pm4.5 $\\ 

\hline
\end{tabular}

\medskip
Notations for channels as in Table~\ref{tab-ns-nod}.
For all GC models, except 47~Tuc and Terzan~5, the numbers 
are scaled per 200 000 $M_\odot$ stellar population mass at the age of 11 Gyr;
for 47 Tuc the numbers are given per its total mass taken as $10^6$ $M_\odot$,
for Terzan~5 - per 370 000 $M_\odot$.
Numbers show the results averaged over all runs for each cluster model (for the number of runs see  Table~\ref{tab-model}).
\end{table*}

\subsection{Effects of dynamics}

The effects of dynamics on the production and retention of NSs are illustrated in Table~\ref{tab-ns-dyn}.
The number of formed CC and EIC NSs is in agreement with the 
retained fraction of the stellar population where binaries have been partially depleted
and some of the SN progenitors were located in the halo, where the escape speed is lower than in the core.
We find that the production of NSs via AIC and MIC is strongly  enhanced 
by dynamical encounters compared to a field population. 
As a result, those formation channels together become comparable to the EIC channel, and produce many more retained NSs than the CC channel. 
A metal-poor cluster, following the trend of field populations, possesses a few per cent more NSs.

As noted above (\S3.1), MIC could lead to the formation of magnetars, which also appear
as anomalous X-ray pulsars (AXPs).
Currently, there are no AXPs detected in old GCs, and only one AXP is detected in 
a young ``super star cluster,'' Westerlund~1 \citep{Muno06}.
On the other hand, all known AXPs are younger than 10$^5$ yr. 
In this case, AXPs could be found in GCs only if their formation rate
is above 1 per 200 (GCs) per 10$^5$ yr, 
or at least 500 per typical cluster, over the cluster lifetime. 
Our formation rates are significantly smaller and therefore 
the hypothesis of a MIC connection to AXPs does not contradict our results.

\subsection{Location}

We consider the position of retained NSs in the cluster.
The progenitors of CC NSs are mainly massive stars with masses $\ga 20 M_\odot$.
However a significant fraction of progenitors of EIC NSs, especially those that evolved from a primordial 
binary, have intermediate initial masses---as low as 4 $M_\odot$. 
The evolutionary time-scales for NSs to be formed are 
$\sim 10^7$ yr for an CC NS and  $\sim 10^{8}$~yr 
for an EIC NS. These are only $\sim ~2$ per cent and  $\sim ~10$ per cent, correspondingly,
of the cluster half-mass relaxation time $t_{\rm rh}$, where 
$t_{\rm rh}=10^9$ yr in a typical cluster.
Even massive stars do not experience strong mass segregation during $\sim0.02 t_{\rm rh}$,
the same is true for stars of intermediate masses that produce EIC NSs 
\cite[see, e.g., Fig.7 in ][, where the mass segregation with time
in a cluster with stars between 0.2 $M_\odot$ and 120  $M_\odot$ is shown]{Ato04_ms}.
When a NS is formed, the mass of the star
is significantly decreased. As a result, a significant fraction of NSs 
is formed in the halo, and many will remain there.
As many  as $\sim 60$ per cent of all EIC NSs in the case of a typical cluster ($t_{\rm rh}=10^9$ yr) 
remain in the halo at 11 Gyr.
AIC and MIC NSs are also present in the halo, but in a lesser proportion, 
as they are mainly produced in a dense environment.
Many of the halo AIC and MIC NSs have recoiled from the core as a result of
dynamical encounters. 
A significant difference with the distribution of NSs over the clusters can only be 
expected for initial mass segregation, or if the initial $t_{\rm rh}$ 
was much shorter than the current $t_{\rm rh}$ resulting in faster mass segregation at early stages. 
Observations find \citep[e.g.][]{Grindlay02} that the radial distribution 
 of MSPs within globular clusters in 47 Tuc is consistent with the production 
 of all MSPs within the core \citep[but cf. ][]{GurkanRasio03}.

\section{Formation of binaries with neutron stars}

\subsection{Primordial binaries with NSs}

\label{ns_prim}

The probability for a NS to remain in a primordial binary
depends strongly on its formation channel (see Table~\ref{tab-ns-bin-prim}).
By definition, MIC leads to the formation of a single NS.
AIC occurs in a very tight binary.  Accordingly, a 
NS formed this way will most likely remain in a binary.
Among CC NSs or EIC NSs,  only several per cent remain in binaries immediately
after NS formation. Binaries that survive SN generally 
have relatively massive secondaries that are also likely to experience SN, 
or the binary may evolve through dynamically unstable mass transfer resulting in a merger,
with stronger binary depletion in the case of initially more massive binaries with a CC NS.
As a result, very few NS binaries formed through those channels remain 
in binaries.
Post-AIC binaries are the GC primordial binaries most likely to contain a NS.
The low probability for CC NSs and EIC NSs to be members of a primordial binary does not change strongly with metallicity.

The formation rate of AIC NSs in GCs shows that as many as 40 NS-MS binaries can be formed in
a typical cluster. Most of them will evolve through the LMXB stage 
before the age of 11 Gyr, producing as a result 20-30 recycled pulsars,
although those MSPs may have short lifetimes and not be observable today (see \S\ref{msp-mild-rec}).
Overall, NSs that remain in binaries may evolve through the mass-transfer phase,
but most such events occur very early in the cluster history, mainly within the first 2 Gyr.  
In more detail, per sample of 270 000 NSs formed in primordial binaries (Z=0.005), 
no binaries started the MT with a MS star after the stellar population reached the age of 8 Gyr;
for  ages between 10 and 11 Gyr we encountered only 1 LMXB with a RG and 2 LMXBs with a WD companion.
From these stellar population runs it is clear that it is highly unlikely that LMXBs observed in GCs of our Galaxy today are primordial. 
Exceptions to this can occur only if a primordial binary had participated in an encounter,
retaining both its companions but changing binary separation and/or eccentricity,
resulting in mass transfer at older cluster ages.
However, simply being in a binary immediately after NS formation 
strongly increases the chance for a NS to participate
in an encounter and, accordingly, to get another, non-primordial, companion.
The fraction of NS binaries with a NS formed via AIC is therefore expected
to be higher than for other NS channels.

The number of double neutron star (DNS) systems formed in the field is very low, 
typically just a few DNSs per 200,000 $M_\odot$ at 11 Gyr. 
For a typical GC this number will be smaller since some of the primordial binaries will be destroyed dynamically
and most DNSs will be ejected. In fact, more than 99 per cent of the first-born NSs in primordial DNS systems are CC NSs, and, accordingly,
have high natal kicks.
The formation of primordial DNSs occurs very early in the GC evolution. 
Most of them are very tight and merge quickly,
leaving very little chance for one to be observed today. As this paper only uses the results
of the NS formation and evolution in the field as a reference point, 
we refer the reader to the complete study of DNS formation channels from primordial binaries, rates,
appearance with time, metallicity dependence and other details provided in Belczynski et al. 2007 (in prep).
We conclude that essentially no primordial DNSs can be present in Galactic GCs at the current time.

\begin{table}
\caption{Binary fractions for NSs in primordial binaries.}
\label{tab-ns-bin-prim}
\begin{tabular}{@{}l  c c c c  c c c}
\hline
Z & \multicolumn{3}{c}{after NS formation } & \multicolumn{3}{c}{at 11 Gyr} \\
& CC & EIC & AIC  & CC & EIC & AIC   \\
\hline
0.0005 &  2.9\% & 3.0\% & 92.3\%  & 0.28\% & 1.38\% & 78.6\%\\
0.001 &   2.8\% & 3.4\% & 93.6\%  & 0.25\% & 1.41\% & 81.8\%\\
0.005 &   2.6\% & 3.4\% & 96.0\%  & 0.18\% & 1.54\% & 88.0\%\\
0.02 &    1.4\% & 4.2\% & 95.2\%  & 0.12\% & 1.49\% & 85.9\%\\
\end{tabular}

\medskip
The fractions of NSs that remain in primordial binaries immediately 
after NS formation (columns $2\div4$) and at 11 Gyr (columns $5\div7$),
for each formation channel.
 
\end{table}

\subsection{Physical collisions}

\label{sec-pc}

\begin{table}
\caption{NS binaries formed in scattering experiments.}
\label{tab-scat}
\begin{tabular}{@{}l  c c c c c c }
\hline
\hline 
 & Z & $\sigma$& \multicolumn{2}{c}{DCE} & \multicolumn{2}{c}{TC}  \\
 &  &km/s & Tot & MT & Tot & MT  \\
\hline
standard & 0.005 & 10 &  3.68\% & 1.75\% & 2.35\% &  1.20\%\\
metal-poor & 0.0005 & 10 &   3.25\% & 1.30\% & 1.93\% & 0.95\%\\
low-$\sigma$ & 0.05 & 5 & 6.80\% & 3.61\% & 9.24\% & 4.37\%\\
\hline
\end{tabular}

\medskip
The table shows the fraction of NSs that successfully formed a binary via physical 
collision with red giants (DCE) or tidal capture (TC). Tot is the total number and MT is the 
fraction of NSs that not only formed a binary, but also started mass transfer
before 11 Gyr.
\end{table}

\begin{figure}
\includegraphics[height=.35\textheight]{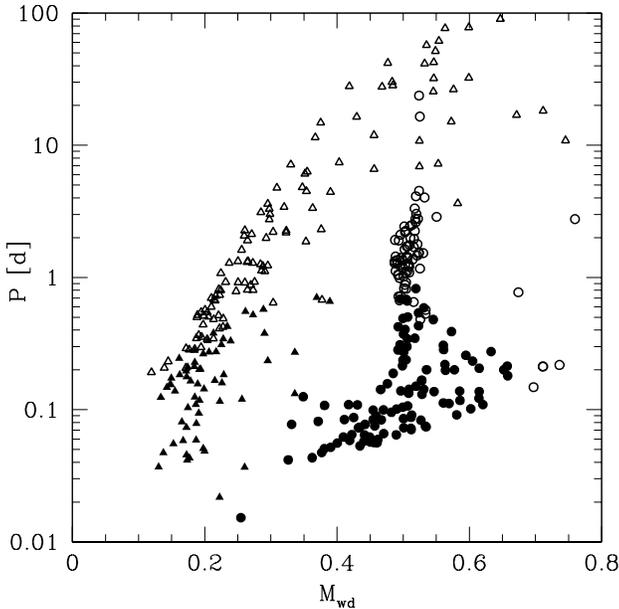}
\caption{
Companions in binary systems formed via physical collision with a red giant 
during the scattering experiment (Z=0.001).
Triangles are WDs, circles are He stars; solid symbols denote the systems that
started the mass tranfer before 11 Gyr.
\label{fig-scat-nswd}
}
\end{figure}

A binary with a NS and a WD can be formed via physical collision with a giant
(see \S\ref{bf_pc}; for earlier works see \citet{Verbunt87_ucxb,Davies92}). 
This binary formation mainly works during single-single star collisions,
although it may occur during  a binary-single encounter as well.
As we showed above, the number of NSs that can be retained in globular clusters
is about a couple hundred for a typical globular cluster.
The cross-section for two stars that have a relative velocity at infinity $v_\infty$
to pass within a distance $r_{\rm max}$ is

\begin{equation}
\sigma = \pi r^2_{\rm min} (1 + \frac{2 G (m_1+m_2)}{r_{\rm max} v^2_\infty}) \ ,
\end{equation}
where the second term is gravitational focussing and in, stellar systems with low velocity dispersion
like GCs, it exceeds the first term.
The time-scale for a NS to  undergo an encounter with a red giant 
can be estimated as $\tau_{\rm coll} = 1/f_{\rm RG}n\sigma v_\infty$, where $f_{\rm RG}$
is the relative fraction of red giants in the stellar population:

\begin{equation}
\tau_{\rm coll} = 1.6\times 10^{11} {\rm \ yr\ } v_{10} n_5^{-1} (M_{\rm NS} + M_{\rm RG})^{-1} R_{\rm max}^{-1}
f_{\rm RG}^{-1} \ ,
\label{eq-tau-coll}
\end{equation}
where $M_{\rm NS}$ and $M_{\rm RG}$ are the masses of the NS and 
 red giant in $M_\odot$, and $R_{\rm max} = r_{\rm max}/R_\odot$
is the maximum distance between the stars during the encounter that leads to the binary formation.  
\cite{Lombardi06_sph} shown that $r_{\rm max}\approx 1.5 R_{\rm RG }$, where $R_{\rm RG }$
is the radius of the red giant.
Averaged over time,  $R_{\rm RG }$ is a few solar radii \citep{Ivanova05_ucxb}, 
and $f_{\rm RG}\approx 5$ per cent.
Using eq.~(\ref{eq-tau-coll}) we estimate that during 11 Gyr about 5 per cent of NSs can participate in the formation of binaries through physical collisions. 

As the estimate shows, having about $200$ NSs in our models of GCs (and about 400 for the models with the highest resolution, $2\times 10^6$ stars), 
we can form a maximum of 10 binary systems through the whole cluster simulation.
To study statistically the characteristics of post-encounter binaries, 
we performed a scattering experiment; 
we considered a population of $10^6$ single stars, 
with an IMF from 0.2$M_\odot$ to 3.0$M_\odot$ and 10000 NSs of $1.4 N_\odot$.
All stars are initially placed in the core with $n_{\rm c}=10^5$ pc$^{-3}$. 
We considered three different models - standard, low metallicity, and with 
low velocity dispersion (see also Table~\ref{tab-scat}).
In these simulations only single-single encounters were allowed.

The scattering experiment shows that 3.7 per cent of all NSs 
successfully formed a binary 
via physical collision with a RG or an AGB star. 
This fraction slightly decreases when the metal-poor population
is considered and increases by a factor of about 2 (in accordance with eq.~\ref{eq-tau-coll})
in the case of lower velocity dispersion. The fraction of NS--WD systems that started 
MT before 11 Gyr is high, $\sim 50$ per cent (see Table~\ref{tab-scat}).

In Fig.~\ref{fig-scat-nswd} we show the masses of companions
in post-collision binaries vs. orbital periods. 
We form about the same number of systems with WD companions and
stripped He star companions (from collisions with AGB stars).
Systems with He star companions are mainly formed within the first 2 Gyr and start  MT
mostly within 3 Gyr. After 3 Gyr both the formation rate and the appearance rate
are slightly higher for systems with WD companions. 
The evolution of mass-transferring systems with He star companions is a bit different from those
with WD companions. They are a few times wider (for the same companion mass) 
and do  not live as long in the mass-transferring phase at high luminosities.

In our numerical simulation of different models of GCs the actual rates are smaller. 
For instance, not all NSs are immediately present in the core
at the time of their formation, and about half of them are still in the halo at the age of 11 Gyr. 
As a result, our ``standard'' model shows about 2 per cent formation rate per NS present in the core
(or 1 per cent per all NSs in the cluster), producing only 2 systems. 
The scatter between several realizations is big. In some simulations we formed only half as many binaries
via this channel, and, in one simulation, 3 times more 
binary systems were formed.
For our standard model, no simulations failed to produce binaries. 
We note that the low-$\sigma$ cluster model, unlike the scattering results, does not produce twice as many  
NS-WD binaries as the standard model -- as the low-$\sigma$ cluster retains less NSs than the standard.

\subsection{Tidal captures}

In our previous study, devoted to the formation of cataclysmic variables (CVs)
in globular clusters \citep{Ivanova06_wd}, we found that tidal capture plays a relatively 
small role in the  formation of dynamical binaries, only $\sim 1-2$ per cent of the total
formation rate.
Two main reasons were responsible: (a) a significant fraction of CVs
have been formed directly from primordial binaries and (b) both physical
collisions and exchange encounters occurring with 
WDs 
form tighter binaries, increasing their chance to start the MT. 
For the case of tidal captures with NSs, the results are different.

In our scattering experiments, 
we find that only $\sim 2$ per cent of NSs form a binary via TC (see Table~\ref{tab-scat});
with a small decrease in the case of a metal-poor population and an  increase (4 times)  
for the lower velocity dispersion case.
The latter effect is due to $r_{\rm max}$ being a 
function of the relative velocity, and increasing as the relative velocity
decreases (see \cite{Ivanova06_wd}).

The rate of tidal captures is flat both with time and companion mass.
In the case of the metal-poor population, 
most TC systems that start MT before 11 Gyr have formed before 5 Gyr. 
For our standard metal-rich population, 
among systems formed in the past few Gyrs, only 
those with MS companions $>0.85 M_\odot$ are able to start MT.
We conclude that  $< 1$ per cent of NSs can form a LMXB via tidal captures, and
that the number of LMXBs mainly depends not on the current cluster properties,
but on the dynamical conditions in the cluster core more than 5 Gyr ago.
For metal-rich clusters, there is an exception for initially more massive MS companions.

In our numerical simulations the formation rate is even smaller, due to the same reasons as in \S\ref{sec-pc}. 
An average probability in our ``standard'' model for a NS to participate in a TC is 0.6 per cent, 
and half that to form an LMXB via this channel.
None the less, even being very small, the TC formation channel for LMXBs is comparable to that of physical collisions
(see for more details \S\ref{sec-lmxbs}).

\subsection{Exchange encounters}

As can be seen from eq.~\ref{eq-tau-coll}, due to a larger $r_{\rm max}$, 
binary encounters are much more likely than single star collisions.
However, compared to the case of physical collisions and TC,
where very tight binaries are formed, a smaller fraction of the successful 
exchange encounters are expected to result in a binary with MT.  
The reason is that when the binary exchange occurs, the binary separation in
the formed post-exchange binary roughly scales with the pre-exchange binary separation as 
the ratio of the new companion mass to the mass of the replaced companion \citep{Heggie96}.
Therefore, when a single NS becomes a member of a binary system at the age of several Gyrs, 
the post-exchange system expands, as the mass of the replaced companion is always less than
the mass of the NS. In addition, when a binary encounter involves 
a very tight binary, which would be likely to start MT soon even in the case of post-exchange expansion, 
a physical collision during the binary encounter is more likely than a simple exchange \citep{Fregeau04_fewbody}.  

Overall, in our standard model $\sim20-25$ per cent of NSs in a cluster (40-50 per cent if we consider only NSs in the core) 
will successfully form  binaries through  exchange encounters.
 $\sim 80$ per cent of post-exchange binaries will have a MS companion, and $\sim 15$ per cent a WD companion.
Only 1.5 per cent of NSs will begin MT with a MS companion,
and only 0.2 per cent with a WD companion. 
The role of binary exchanges in the formation
of NS-WD LMXBs is therefore negligible compared to physical collisions.

\section{Low-mass X-ray binaries}

\label{sec-lmxbs}

\begin{table*}
\caption{Average appearance rate of LMXBs formed via different dynamical channels.}
\label{tab-lmxb-prod}
\begin{tabular}{@{}l  c c c c c }
\hline
\hline 
 model & AIC  & TC & DCE & BE/MS & BE/WD  \\
\hline
standard & $1.49\pm0.54 $& $0.07\pm^{0.10}_{0.07} $& $0.14\pm^{0.17}_{0.14} $& $0.51\pm^{0.58}_{0.51} $& $0.06\pm^{0.10}_{0.06} $\\ 
metal-poor & $2.40\pm0.54 $& $0.11\pm^{0.15}_{0.11} $& $0.06\pm^{0.13}_{0.06} $& $0.67\pm^{0.73}_{0.67} $& $0.0$\\ 
high-den & $3.62\pm1.42 $& $0.97\pm0.76 $& $1.27\pm0.66 $& $1.81\pm0.71 $& $0.0$\\ 
med-den & $0.54\pm0.19 $& $0.11\pm^{0.15}_{0.11} $& $0.0$& $0.11\pm^{0.15}_{0.11} $& $0.0$\\ 
low-den & $0.22\pm0.23 $& $0.0$& $0.0$& $0.0$& $0.0$\\ 
low-$\sigma$ & $0.77\pm0.44 $& $0.06\pm^{0.13}_{0.06} $& $0.0$& $0.47\pm0.27 $& $0.06\pm^{0.13}_{0.06} $\\ 
long-$t_{\rm rh}$ & $0.87\pm0.68 $& $0.0$& $0.0$& $0.11\pm^{0.15}_{0.11} $& $0.05\pm^{0.12}_{0.05} $\\ 
BF05 & $1.20\pm0.52 $& $0.05\pm^{0.11}_{0.05} $& $0.10\pm^{0.22}_{0.10} $& $0.40\pm0.14 $& $0.0$\\ 
fast-MB & $1.89\pm0.36 $& $0.06\pm^{0.13}_{0.06} $& $0.17\pm0.15 $& $0.89\pm0.36 $& $0.0$\\ 
CE-reduced & $1.32\pm0.53 $& $0.0$& $0.0$& $0.50\pm0.23 $& $0.0$\\ 
oldkicks & $0.0$& $0.0$& $0.0$& $0.11\pm^{0.15}_{0.11} $& $0.0$\\ 
47 Tuc & $8.03\pm1.16 $& $0.28\pm^{0.63}_{0.28} $& $0.56\pm^{0.59}_{0.56} $& $2.49\pm0.36 $& $0.14\pm^{0.31}_{0.14} $\\ 
Terzan 5 & $9.35\pm1.20 $& $1.43\pm0.49 $& $2.21\pm0.42 $& $2.21\pm0.70 $& $0.13\pm^{0.29}_{0.13} $\\ 

\hline
\end{tabular}

\medskip
Number of LMXBs that appear per Gyr at age $11\pm1.5$ Gyr, 
for different formation channels: 
AIC in a dynamically formed WD-binary, 
TC -- tidal capture, 
DCE -- a dynamical common envelope event (physical collision with a red giant), 
BE/MS -- binary exchanges when a MS companion is acquired, 
BE/WD -- binary exchanges when a WD companion is acquired.
\end{table*}

\begin{table}
\caption{Average number of LMXBs with different donors}
\label{tab-lmxb-probability}
\begin{tabular}{@{}l  l l l l }
\hline
\hline 
 model & MS  & RG & WD  \\
\hline
standard & $1.85\pm0.56 $& $0.12\pm^{0.16}_{0.12} $& $0.46\pm0.41 $\\ 
metal-poor & $1.22\pm1.05 $& $0.10\pm0.07 $& $0.16\pm0.16 $\\ 
high-den & $7.31\pm2.21 $& $0.20\pm^{0.27}_{0.20} $& $6.38\pm3.60 $\\ 
med-den & $0.46\pm0.40 $& $0.03\pm^{0.04}_{0.03} $& $0.08\pm^{0.12}_{0.08} $\\ 
low-den & $0.51\pm^{0.63}_{0.51} $& $0.01\pm0.01 $& $0.11\pm^{0.25}_{0.11} $\\ 
low-$\sigma$ & $3.13\pm1.25 $& $0.07\pm0.06 $& $0.12\pm0.11 $\\ 
long-$t_{\rm rh}$ & $0.84\pm0.73 $& $0.09\pm^{0.18}_{0.09} $& $0.11\pm^{0.16}_{0.11} $\\ 
BF05 & $1.12\pm0.32 $& $0.08\pm^{0.12}_{0.08} $& $0.44\pm0.33 $\\ 
fast-MB & $1.38\pm0.93 $& $0.10\pm^{0.11}_{0.10} $& $1.09\pm0.75 $\\ 
CE-reduced & $1.98\pm1.33 $& $0.06\pm0.06 $& $0.23\pm0.16 $\\ 
oldkicks & $0.29\pm^{0.49}_{0.29} $& $0.0$& $0.0$\\ 
47 Tuc & $9.28\pm4.10 $& $0.13\pm0.09 $& $4.57\pm3.03 $\\ 
Terzan 5 & $7.16\pm1.95 $& $0.14\pm0.11 $& $7.73\pm3.03 $\\ 

\hline
\end{tabular}

\medskip
Average number of LMXBs present at ages $11\pm1.5$ Gyr, 
for different donors (RG includes also subgiants and Hertzsprung gap stars).
\end{table}

From the discussion above, one can see that the expected formation rate of LMXBs is rather small.
For all dynamical formation channels that involve a NS directly, in a typical dense metal-rich cluster, 
the combined rate is only  $\sim 3$ per cent per NS. 
This is 6 LMXBs per 11 Gyr of the cluster life, where
2 LMXBs are with a WD companion and 4 LMXBs are with a MS or a RG companion.
A small number of LMXBs may come from primordial, but dynamically modified binaries
(where the encounters changed the binary eccentricity or separation).
In addition, however, some fraction of LMXBs is expected to be provided by dynamically formed binaries with heavy WDs (see \cite{Ivanova06_wd}). 
Such binaries, if evolved through AIC, create binaries that reinstate MT on a NS soon after AIC, with a MS, WD or RG donor. 
From our current simulations, we find that the channel of LMXB production 
from AIC of dynamically formed WD binaries produces more LMXBs than any of  
the other dynamical LMXB production channels discussed above.
Overall, in a typical cluster it is 2-3 times more efficient during 9.5-12.5 Gyr than physical collisions,
tidal captures and binary exchanges (see Table~\ref{tab-lmxb-prod}).
The exceptions are only very high density clusters.
We also note that if the majority of LMXBs are produced through AIC, but these binaries 
do not produce long-lived MSPs (\S\ref{msp-mild-rec}), then the production rates of LMXBs and MSPs 
may not be closely coupled.  However, LMXBs produced by AIC may also have shorter 
lifetimes due to the exhaustion of the companion's mass during the pre-AIC MT stage.

\begin{figure*}
\includegraphics[height=.35\textheight]{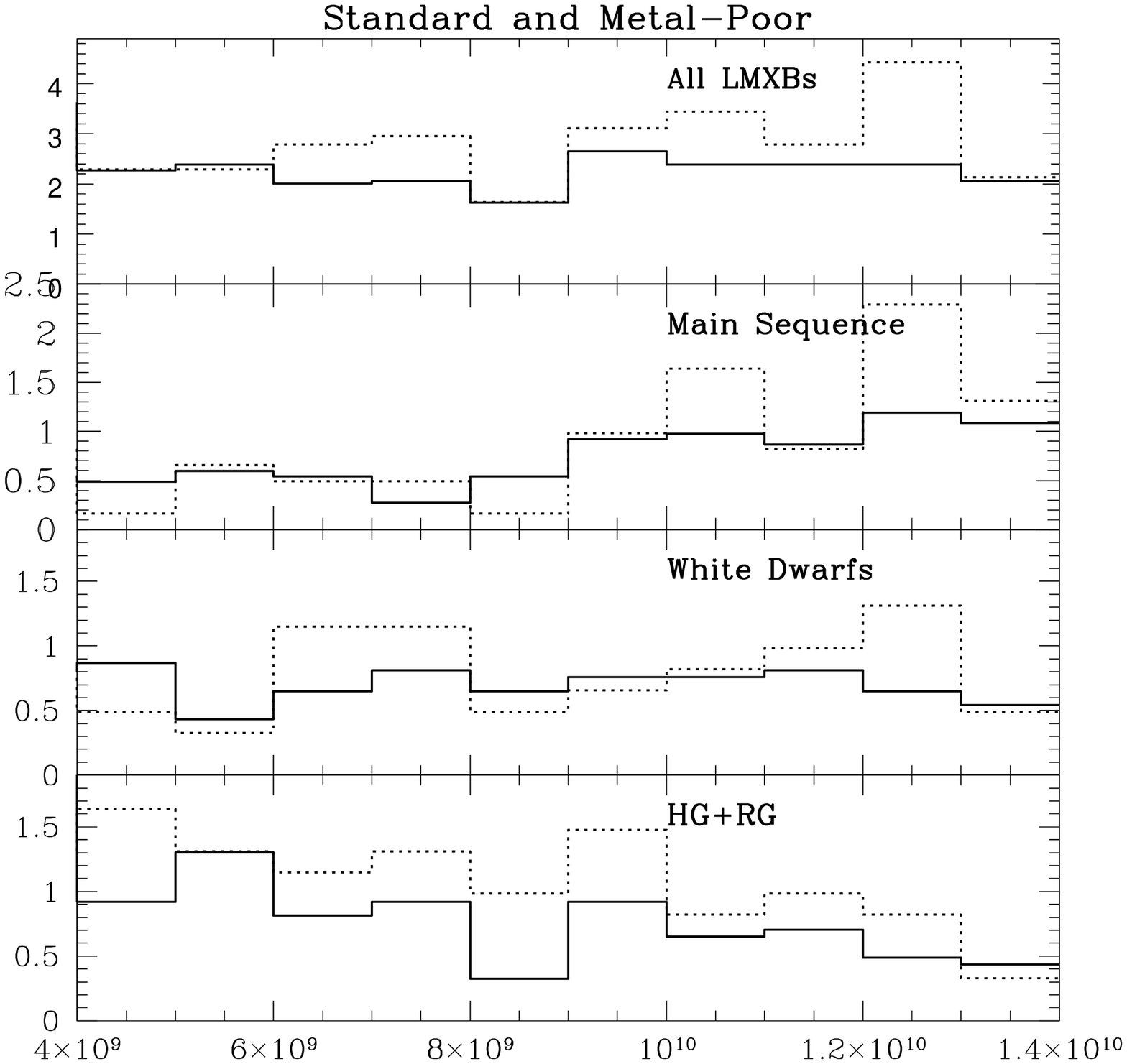} 
\includegraphics[height=.35\textheight]{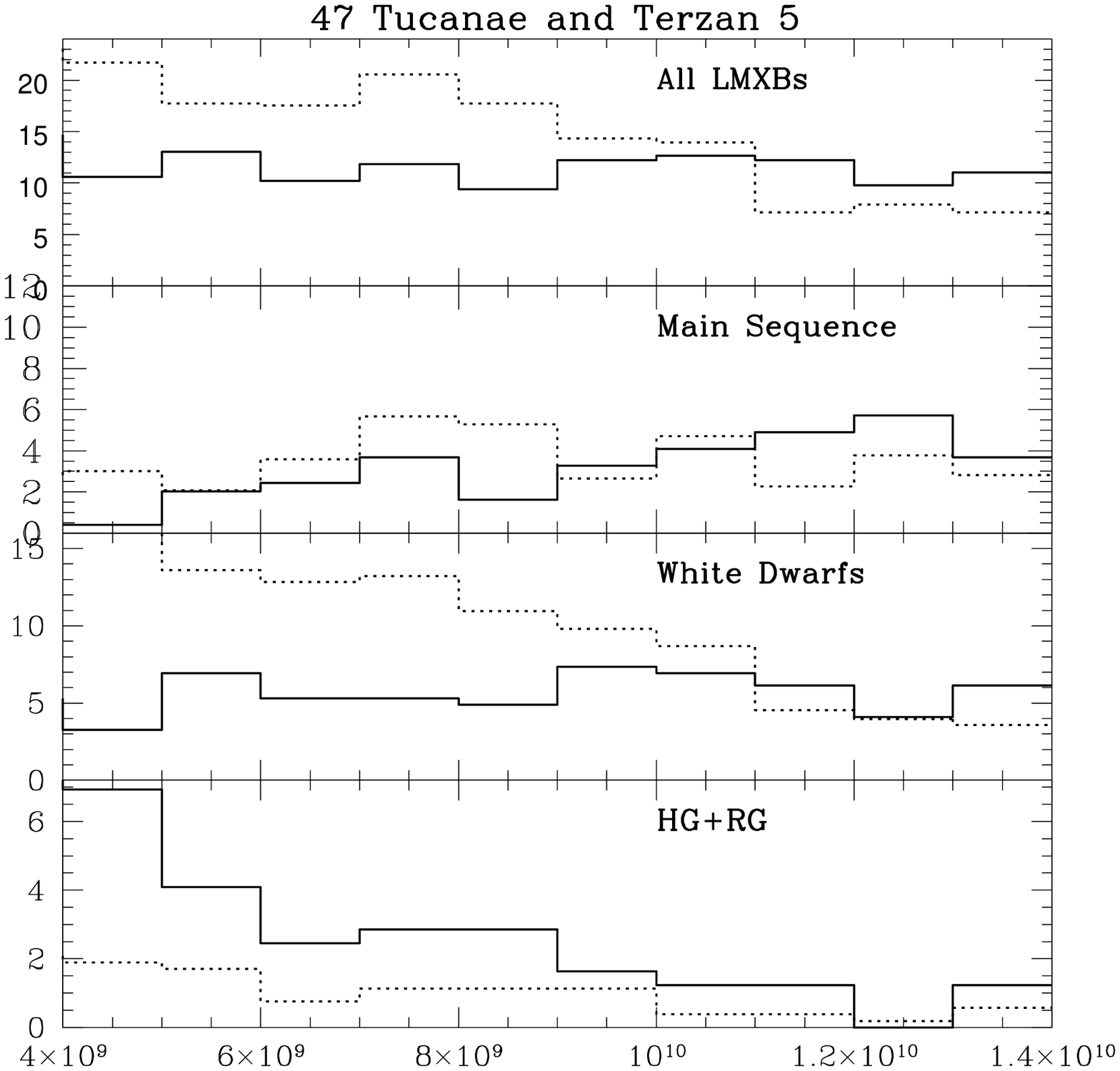}
\caption{
Number of appearing LMXBs per Gyr with different donor types (with all donors, with MS donors, with WD donors and with 
donors that are RGs, subgiants or are in the Hertzsprung Gap). The left panel shows the results for the standard model (solid line) and the 
metal-poor model (dotted line). The right panel shows the results for 47 Tuc (solid line) and Terzan 5 (dotted line).
\label{fig-lmxb-zoo}
}
\end{figure*}

In Fig.~\ref{fig-lmxb-zoo} we show when LMXBs (with different companions) appear during the cluster evolution.
We show only the cluster ages after 4 Gyr, as before that time mainly primordial LMXBs appear.
The connection between the production rates of LMXBs and the numbers that
are present at a particular time in a GC is provided by the life-time $\tau_{\rm LMXB}$ of LMXBs with various donors.
An average $\tau_{\rm LMXB}$ for NS-MS LMXBs is about 1 Gyr. Depending on the metallicity and the initial donor mass,
a system can be persistent 5-40 percent of the MT time for metal-rich donors $>0.6 M_\odot$ and
transient during the rest;
for donors of lower metallicities or smaller masses, a NS-MS LMXB will be transient all the time \citep{Ivanova06_lmxb},
and therefore more likely seen as a qLMXB rather than as a bright LMXB.
An LMXB with a red giant companion or a companion in the Hertzsprung gap is very short-lived,
$10^5-10^7$ yr, and in only very rare cases can they live as long as $10^8$ yr.
In the case of NS-WD LMXBs -- ultra-compact X-ray binaries (UCXBs) -- the total $\tau_{\rm LMXB}$ is a few Gyr,
however the time during which the system is persistent and has an X-ray luminosity above $10^{36}$ erg/s is only $\sim10^8$~yr. 
UCXBs at $\sim10^8$ years will fall below $L_{b\rm ol}=1.3\times10^{36}$ ergs/s.  
According to the theory of irradiated disks \citep{Dubus99_accr} and the evolution 
of UCXBs \citep{Deloye03_wd}, $10^{36}$ ergs/s is roughly the critical $L_X$ 
for maintaining persistent accretion.  \citet{intZand07}  present a table of 
40 persistent X-ray bursters, which shows a striking low-$L_X$ pileup and cutoff 
at $L_X=1.3\times10^{36}$ ergs/s, or 0.5\% of the hydrogen-poor $L_{Edd}$ 
(incorporating their relativistic correction).  
Incompleteness in Galactic surveys does not fully explain this cutoff, 
as the PCA bulge scans monitor many sources at half this flux.  
Therefore we adopt $L_X=1.3\times10^{36}$ ergs/s as the break between 
persistent and transient behavior for UCXBs.

Combining the number of appearing LMXBs with their lifetimes in the simulations, 
we find that a typical dense cluster at $11\pm1.5$ Gyr age 
can contain up to 2 LMXBs with a MS companion and up to 1 LMXB with a WD companion 
(see also Table~\ref{tab-lmxb-probability}, note the very large scatter).
We specify, that Table~\ref{tab-lmxb-probability} shows the average expected number of LMXBs --
both bright LMXBs and qLMXBs.
We choose a cutoff period of 55 minutes for identifying UCXBs as LMXBs.  
We justify this by noting that the longest period UCXB yet identified has a period of 55 minutes 
\citep{Krimm07}, and that the evolution of UCXBs will produce very large numbers 
of longer-period systems which have not yet been seen.
As expected, the number of LMXBs shows a strong dependence on the core density;  
 dependence on the velocity dispersion and the half-mass relaxation time is also observed.

A strong dependence of LMXB production on cluster density has been inferred observationally, 
both in Galactic globular clusters \citep[e.g.,][]{Verbunt87,Pooley03,Heinke03a} 
and in extragalactic globular clusters \citep[e.g.,][]{Jordan04,Sivakof07_lmxb}.
The interpretation has been that $N_{\rm LMXB}$ scales roughly with the encounter frequency 
$\Gamma=\rho_{\rm c}^2 r_{\rm c}^3/\sigma$, where $r_c$ is the core radius \citep{Verbunt87}.

The observations of LMXBs in our galaxy suffer from low number statistics--only 12 
clusters have LMXBs--so attempts to understand the LMXB formation rate from them 
encounter difficulties \citep[e.g.][]{Bregman06}.
Observations of qLMXBs in our galaxy also suffer from low number statistics, 
as only $\sim$two dozen clusters have been observed with Chandra sufficiently 
to identify most qLMXBs. \cite{Pooley06}  and \cite{Heinke06} 
study overlapping samples of clusters using different assumptions.  
Their estimates of the density dependence of qLMXB production for the case 
of no metallicity dependence also overlap, suggesting a density dependence 
larger than suggested by $\Gamma$, but with large errors. 
\cite{Heinke06} also investigate (inconclusively) 
the possibility of metallicity dependence, 
which decreases the required density dependence.

Observations of LMXBs in globular clusters of other galaxies do not suffer 
from low statistics, and thus have clearly demonstrated a strong dependence 
of LMXB production on metallicity \citep{Kundu02,Kundu07}.  
Analyses by \cite{Jordan04} and \cite{Sivakof07_lmxb} both find that 
LMXB production has a density dependence lower than $\Gamma$.  
However, it is extremely difficult 
to determine the structural properties of 
extragalactic globular clusters with high accuracy, so uncertainty in the density dependence remains.

\begin{figure}
\includegraphics[height=.35\textheight]{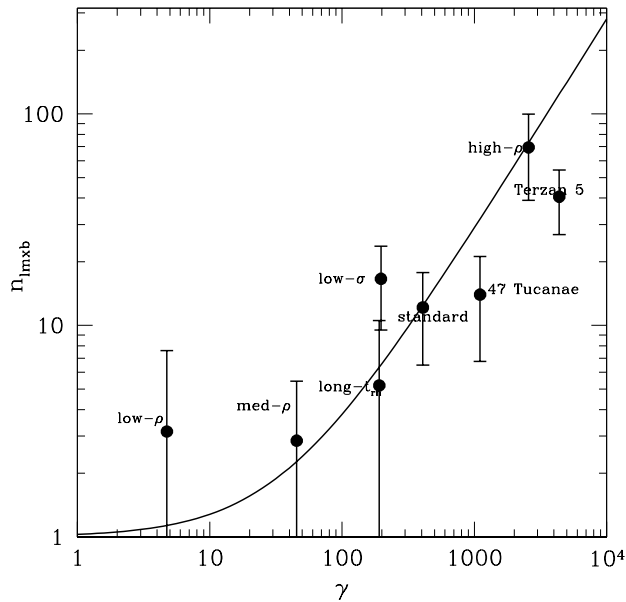}
\caption{
The specific collision numbers $\gamma$ and numbers of LMXBs in simulated clusters.
The solid line corresponds to $n_{\rm LMXB} = (1^{+7}_{-1}) +0.028\gamma$.
The error bars correspond to the scatter in our simulations.
\label{fig-gamma}
}
\end{figure}

In Fig.~\ref{fig-gamma} we demonstrate how the specific number of 
LMXBs $n_{\rm LMXB}=N_{\rm LMXB}/M$ depends on the specific collision frequency $\gamma=\Gamma/M$,
taken per units of $10^6\ M_\odot$, as in \cite{Pooley06}.
To calculate $\Gamma$ in the same way as observers, we extracted values of $\rho_c$ and 
$r_{\rm c}$ from our models at the cluster age of 11 Gyr, and used our input parameter $\sigma$.
 When only cluster density is varied, 
$n_{\rm LMXB}$ depends linearly on $\Gamma$, however all the other parameters create significant scatter. 
We therefore urge not to overinterpret the dependence of the number of observed LMXBs on $\Gamma$, 
or the meaning of the power-law dependence of $\Gamma$ on density; the scatter in $\Gamma$ 
in the observed GCs could easily be explained by the variance in other dynamical properties. 
If one assumes that in the observed sample of GCs the density increase is likely to be accompanied by increases
of the cluster mass, and, accordingly, $t_{\rm rh}$ and $\sigma$, then we expect from our results that 
the power law dependence of $n_{\rm LMXB}$ with $\gamma$ is a bit less than one and close to the
value of 0.8, as in \cite{Sivakof07_lmxb}, where metallicity effects are separated from 
dynamical effects.

Even though a typical cluster contains 1 or 2 NS-MS LMXBs, 
there is only $\sim 10-20$ per cent chance that those LMXBs will be bright and persistent, and then only for metal-rich clusters.
The probability to contain a bright persistent UCXB is no more than 10 per cent and does not depend strongly on the metallicity.
Overall, there is at best a 30 per cent chance to contain a bright LMXB in a typical metal-rich cluster and a 10 per cent chance in a metal-poor cluster.
This may explain the observed ratio between the bright LMXBs in metal-poor and metal-rich clusters. However, to confirm that, 
it is necessary to have observational statistics on qLMXBs in both metal-poor and metal-rich clusters -- 
if the explanation above is correct, the number of qLMXBs will not show a dependence on the metallicity.
Considering rich Galactic globular clusters, we find that our formation rate of UCXBs at 11 Gyr in Terzan 5 is about 
10 per Gyr and in 47~Tuc, $\sim 5$ per Gyr (see Fig.~\ref{fig-lmxb-zoo}).
Since  $\tau_{\rm UCXB}\sim10^8$ yr, this  agrees with the presence of one bright (possibly ultracompact) LMXB in Terzan 5 
and no bright UCXB detected in 47 Tuc.

To compare the entire population of UCXBs in GCs with the results of our simulations, we integrated through the whole
sample of galactic globular clusters, using data for velocities from 
\cite{Gnedin02}\footnote{see on-line data at http://www.astro.lsa.umich.edu/\~ognedin/gc/vesc.dat} and for other quantities from
\cite{Harris96}. 
For this calculation, we used the formation rates shown in Table~\ref{tab-lmxb-prod}
and assumed that an UCXB will be bright for only $\sim 10^8$ yr. 
As a result, the integral value of bright UCXBs in our Galaxy is 7.5,
which is pretty close to the observed number of 4 to 9 UCXBs in GCs. 
The clusters that are most likely to contain UCXBs overlap very well with the clusters where those UCXBs 
are observed: NGC 1851 and NGC 7078 each have a probability to contain an UCXB $>$ 20 per cent.  
NGC 6712 is known to have been heavily tidally stripped, so its current cluster conditions 
do not reflect the conditions that produced its UCXB  \citep[see discussion in][]{Ivanova05_ucxb}.  
Core-collapsed clusters are not properly modeled by our code, 
so our failure to predict UCXBs in NGC 6624 and NGC 6652 (a likely UCXB) is not meaningful.  
Other clusters that have a $>$ 20 per cent chance to contain a UCXB include 
47 Tuc, Terzan 5, NGC 6266, NGC 6388, NGC 6440, NGC 6441, NGC 6517, and NGC 6715 (M54).
Unsurprisingly, this list consist of clusters with a high collision number.  
Those clusters have not been shown to contain bright UCXBs, 
but they generally contain numerous quiescent LMXBs (some of which may be UCXBs) 
and MSPs (some of which may be products of UCXB evolution).  Overall, it seems
that physical collisions are able to produce UCXBs in numbers comparable to those observed in globular clusters in the Milky Way.

It is more difficult to make a similar comparison for NS-MS LMXBs, as the lifetime at the bright stage is much less certain.
In the whole GC population we expect to have about 180 qLMXBs formed through tidal captures and binary exchanges.
We note that since our numbers for core-collapse clusters are lower limits, the real number of qLMXBs is expected to be higher.
We identified those clusters that are predicted to contain 5 or more NS-MS LMXBs, most of which will be in quiescence. 
Those clusters are the same 10 as noted above for UCXBs:  
47 Tuc, Terzan 5, NGC 1851, NGC 6266, NGC 6388, NGC 6440, NGC 6441, NGC 6517, M 54, and NGC 7078. 
5 of them, 47 Tuc, Terzan 5, NGC 6266, NGC 6388 and NGC 6440 are known to contain 5 or more qLMXBs
\citep[][Cohn et al. 2008 in prep, ]{Heinke03a}.
NGC 6441 and NGC 6440 contain bright LMXBs with MS donors \citep{Verbunt06,Altamirano07}.
We cannot make clear predictions for the core-collapsed, 
heavily reddened clusters Terzan 1, Liller 1, and Terzan 6 because 
we do not model dynamical evolution of clusters experiencing core collapse.  
We note that if AIC binaries contribute to LMXB formation, the numbers will go up, although not strongly
as the characteristic MT time in this case is shorter than in the case of an average dynamically formed NS-MS binary.
According to our models,
several clusters have probability 0.3 or more to contain a RG-NS LMXB, one of which -- NGC 7078  -- 
indeed holds one.

For extragalactic globular clusters, \citep{Bildsten04_ucxb} showed that the observed luminosity functions can be explained
by a constant birthrate of UCXB binaries, 10 per Gyr per 200,000 $M_\odot$, for all GCs in a galaxy. 
We obtained almost constant birthrates for such binaries, but our rates are significantly lower 
(one per Gyr for  200,000 $M_\odot$ in a typical ``dense'' cluster). 
In fact, a 10 times higher UCXB formation rate 
will produce 
as many as 10 bright persistent LMXBs in Terzan 5.
It is possible that the typical density of GCs in ellipticals is larger than that of GCs in our galaxy,
which may explain the variation in LMXB production.
However, since we do not find any difference in UCXB production between metal-poor and metal-rich populations,
our simulations do not support the idea that most X-ray binaries in extragalactic GCs are UCXBs.
To explain most X-ray binaries in extragalactic GCs with UCXBs, we would need to assume that more 
NSs are retained (or formed) in metal-rich than metal-poor clusters 
(since the model of an irradiation-induced wind from \cite{Maccarone04_metal} is not applicable for degenerate donors).

\begin{table*}
\caption{Plausible pulsars and their location at 11 Gyr.}
\label{tab-puls-dyn}
\begin{tabular}{@{}l c c c c c c c c c}
\hline
model & \multicolumn{3}{c}{Halo} & \multicolumn{3}{c}{Core} & \multicolumn{3}{c}{Ejected} \\
& All psrs & Bin psrs  & NS-bin & All psrs & Bin psrs  & NS-bin & All psrs & Bin psrs  & NS-bin   \\
\hline
standard & $25.3\pm3.4 $ & $17.4\pm3.3 $ & $19.5\pm3.1 $ & $39.6\pm4.3 $ & $17.2\pm2.7 $ & $19.9\pm2.8 $ & $ 105\pm9.6 $ & $36.5\pm5.8 $ & $41.6\pm6.2 $ \\ 
metal-poor & $24.2\pm4.8 $ & $18.4\pm4.0 $ & $20.7\pm4.1 $ & $43.7\pm5.6 $ & $19.9\pm2.1 $ & $23.3\pm2.6 $ & $ 128\pm 11 $ & $37.2\pm5.1 $ & $51.9\pm 11 $ \\ 
high-den & $42.4\pm9.9 $ & $32.0\pm5.0 $ & $35.3\pm5.5 $ & $98.9\pm5.4 $ & $22.5\pm4.6 $ & $23.2\pm4.6 $ & $ 228\pm5.9 $ & $ 142\pm 12 $ & $ 159\pm 16 $ \\ 
med-den & $11.8\pm3.4 $ & $11.6\pm3.7 $ & $15.0\pm3.3 $ & $16.2\pm2.9 $ & $14.3\pm3.2 $ & $19.0\pm4.3 $ & $26.6\pm2.3 $ & $26.3\pm2.3 $ & $62.5\pm 36 $ \\ 
low-den & $22.3\pm1.4 $ & $15.7\pm1.8 $ & $17.7\pm2.4 $ & $19.2\pm3.0 $ & $13.9\pm1.8 $ & $17.3\pm2.2 $ & $86.3\pm9.5 $ & $26.1\pm5.9 $ & $42.9\pm 15 $ \\ 
low-$\sigma$ & $ 9.1\pm3.7 $ & $ 5.3\pm2.1 $ & $ 6.4\pm1.8 $ & $14.4\pm2.9 $ & $ 3.6\pm1.7 $ & $ 4.6\pm2.3 $ & $ 170\pm 10 $ & $88.7\pm8.6 $ & $ 106\pm 14 $ \\ 
long-$t_{\rm rh}$ & $31.3\pm5.2 $ & $22.7\pm4.4 $ & $25.9\pm5.0 $ & $17.3\pm2.1 $ & $ 7.7\pm3.3 $ & $ 8.6\pm3.2 $ & $94.0\pm9.9 $ & $30.8\pm4.1 $ & $55.8\pm 24 $ \\ 
BF05 & $15.7\pm3.8 $ & $10.4\pm3.4 $ & $12.4\pm2.9 $ & $29.9\pm5.9 $ & $13.9\pm2.6 $ & $15.8\pm2.6 $ & $68.6\pm7.9 $ & $24.4\pm5.0 $ & $38.0\pm 13 $ \\ 
fast-MB & $25.0\pm3.3 $ & $16.2\pm3.5 $ & $18.3\pm4.3 $ & $46.4\pm5.0 $ & $19.3\pm2.7 $ & $22.2\pm2.5 $ & $ 110\pm 11 $ & $39.2\pm5.1 $ & $50.3\pm9.9 $ \\ 
CE-reduced & $ 7.8\pm1.1 $ & $ 3.2\pm1.5 $ & $ 4.9\pm2.5 $ & $30.8\pm3.2 $ & $12.0\pm2.4 $ & $15.3\pm2.9 $ & $17.7\pm1.0 $ & $ 7.5\pm2.3 $ & $18.7\pm 11 $ \\ 
oldkicks & $ 8.9\pm3.7 $ & $ 0.3\pm^{0.7}_{0.3} $ & $ 0.3\pm^{0.7}_{0.3} $ & $10.2\pm3.2 $ & $ 2.1\pm1.6 $ & $ 2.3\pm1.8 $ & $67.0\pm4.7 $ & $ 4.4\pm2.3 $ & $12.4\pm4.6 $ \\ 
47 Tuc & $ 146\pm 15 $ & $84.8\pm4.7 $ & $ 104\pm9.1 $ & $ 181\pm 16 $ & $77.1\pm 11 $ & $85.6\pm 15 $ & $ 409\pm 18 $ & $ 111\pm 13 $ & $ 136\pm 11 $ \\ 
Terzan 5 & $56.1\pm7.7 $ & $38.9\pm3.7 $ & $40.3\pm2.4 $ & $ 203\pm 12 $ & $58.4\pm10.0 $ & $64.1\pm9.1 $ & $ 228\pm 12 $ & $ 142\pm 10 $ & $ 178\pm 32 $ \\ 

\hline
\end{tabular}
\medskip

The number of ``plausible'' pulsars  that are retained in the halo or core of the cluster, as well as the number of ejected pulsars.
As a pulsar we define here a NS that gained mass $>0.01 M_\odot$ after its formation.
``All psrs'' - the number of all pulsars, ``Bin psrs'' 
- the number of binary pulsars and ``NS-bin'' - the number of all NSs in binaries, provided for comparison.
For all GC models, except 47~Tuc and Terzan~5, the numbers 
are scaled per 200 000 $M_\odot$ stellar population mass at the age of 11 Gyr;
for 47 Tuc the numbers are given per its total mass taken as $10^6$ $M_\odot$,
for Terzan~5 - per 370 000 $M_\odot$.
\end{table*}

\section{Millisecond Pulsars}

\subsection{{Populations of NSs that have gained mass}}

\label{ns-gain}

In Table~\ref{tab-puls-dyn} we show the number of plausible pulsars that are formed in our simulations.
In this table, we count  as pulsars (i) all NSs that gained more than 0.01 $M_\odot$ after their formation,
through all the possible mechanisms of mass gain (mass transfer, common envelope accretion and 
during mergers or physical collisions) -- including all recycled pulsars (see also \S\ref{msp-scenario});
(ii) all NSs that were formed through AIC/MIC within the last $10^8$ years -- ``slow'' high-$B$ field pulsars 
(see also \S\ref{msp-nonrec}).
To investigate the role of mildly recycled post-AIC and post-collisional NSs, we include all of them in  
Table~\ref{tab-puls-dyn} 
(c.f. \S\ref{msp-mild-rec} and  \S\ref{msp-fast-rec}), as well as 
all UCXBs that have evolved past 55 minutes 
(c.f. \S\ref{mt-bmsp});
the reduction of the simulated populations according to lifetime criteria will be done later in this Section.
Three important facts can be seen from this table: 
(i) the numbers of ejected and retained pulsars are comparable;
(ii) almost all NSs in binaries are pulsars; 
(iii) the numbers of produced pulsars are very large, more than several times the numbers of detected pulsars in specific comparable clusters 
(compare 320 MSPs in simulations vs. 22 detected MSPs for 47 Tuc, or
250 MSPs in simulations vs. 33 detected MSPs in Terzan 5).

The last point shows that a direct comparison of the numbers of NSs that gained mass with the observed 
numbers of radio MSPs fails, unless the detection probability is only about 10 per cent  
\citep[however this should be at least 33 per cent and may be even close to one, see][]{Heinke05a}. 
The alternative is that we produce and retain in our simulations rather more NSs 
than are retained in real GCs.
In this case we cannot explain with our simulations the observed number of LMXBs.  
We conclude therefore that the number of NSs that gain mass after their formation is
correct, but, as was discussed in \S\ref{msp-scenario}, 
not all the mechanisms for mass gain lead to the formation of a radio MSP.

Let us analyze the population of NSs which have gained mass in more detail.
In Tables~\ref{tab-puls-zoo-ter} and \ref{tab-puls-zoo-tuc} 
we show the results of simulations of 47 Tucanae and Terzan 5, distinguishing NSs with gained mass by their formation mechanism 
and by the process through which they acquired mass after the NS formation.
Note that the plausible pulsars that were formed via AIC or MIC constitute more than half of all NSs that gained mass after their formation. 
As we discussed in \S\ref{msp-mild-rec}, such NSs may not reduce their magnetic field 
or may be failed to spun up and accordingly could have only very short
life-times as a detectable MSP (``2a'', ``2b'').
Indeed, very few of them were recycled recently, within last $10^8$ years
(see Tables~\ref{tab-puls-zoo-ter} and \ref{tab-puls-zoo-tuc}).
From this moment, we remove from our further consideration of the 
population of ``present'' radio pulsars those AIC NSs that
were recycled or formed more than $10^8$ yr ago (see also \S\ref{msp-nonrec}, \S\ref{msp-mild-rec}).  

We produce MSPs in AIC systems  more efficiently than \cite{Kuranov06}.
The most plausible explanation is the different adopted natal kicks distributions for ECS NSs, 
as they adopted a Maxwellian distribution with RMS 250 km/s (cf. 40 km/s in our case). 
Nonetheless, we we find that 
AIC evolution does not lead to the formation of long-living radio MSP
and therefore can not explain most of the population of MSPs in GCs (c.f.  \cite{Kuranov06}).

\subsection{Constraints from comparison with the observed populations of bMSPs}

\begin{table}
\caption{Types of pulsars, at 11 Gyr, Terzan 5.}
\label{tab-puls-zoo-ter}
\begin{tabular}{@{}l c c c c }
\hline
 & \multicolumn{2}{c}{Halo} & \multicolumn{2}{c}{Core} \\
& Total & Binary  & Total & Binary \\
\hline
\\
\multicolumn{5}{l}{All ``plausible''  pulsars} \\ 
All & $56.1\pm7.7 $ & $38.9\pm3.7 $ & $ 203\pm 12 $ & $58.6\pm 10 $ \\ 
MT & $ 7.7\pm2.4 $ & $ 7.4\pm2.2 $ & $22.3\pm2.9 $ & $10.8\pm2.7 $ \\ 
MT-WD & $29.5\pm3.1 $ & $29.1\pm3.4 $ & $57.1\pm 12 $ & $37.2\pm 10 $ \\ 
CE & $ 1.9\pm1.8 $ & $ 1.7\pm^{1.8}_{1.7} $ & $ 6.2\pm3.1 $ & $ 1.5\pm1.1 $ \\ 
DCE & $ 0.0\pm0.0 $ & $ 0.0\pm0.0 $ & $ 2.6\pm1.8 $ & $ 2.6\pm1.8 $ \\ 
Merg & $17.0\pm4.1 $ & $ 0.8\pm^{0.8}_{0.8} $ & $ 115\pm6.4 $ & $ 6.4\pm1.8 $ \\ 
\multicolumn{5}{l}{Core Collapse} \\ 
All & $ 6.6\pm0.9 $ & $ 0.0\pm0.0 $ & $ 4.5\pm1.4 $ & $ 0.8\pm0.4 $ \\ 
MT & $ 0.0\pm0.0 $ & $ 0.0\pm0.0 $ & $ 0.2\pm^{0.4}_{0.2} $ & $ 0.0\pm0.0 $ \\ 
MT-WD & $ 0.0\pm0.0 $ & $ 0.0\pm0.0 $ & $ 0.8\pm^{0.8}_{0.8} $ & $ 0.2\pm^{0.4}_{0.2} $ \\ 
CE & $ 0.0\pm0.0 $ & $ 0.0\pm0.0 $ & $ 0.0\pm0.0 $ & $ 0.0\pm0.0 $ \\ 
DCE & $ 0.0\pm0.0 $ & $ 0.0\pm0.0 $ & $ 0.2\pm^{0.4}_{0.2} $ & $ 0.2\pm^{0.4}_{0.2} $ \\ 
Merg & $ 6.6\pm0.9 $ & $ 0.0\pm0.0 $ & $ 3.4\pm1.6 $ & $ 0.4\pm^{0.5}_{0.4} $ \\ 
\multicolumn{5}{l}{ECS} \\ 
All & $10.4\pm1.8 $ & $ 3.8\pm1.6 $ & $80.9\pm9.1 $ & $11.9\pm3.0 $ \\ 
MT & $ 1.1\pm0.8 $ & $ 0.9\pm0.7 $ & $ 8.3\pm2.7 $ & $ 2.8\pm2.1 $ \\ 
MT-WD & $ 0.4\pm^{0.5}_{0.4} $ & $ 0.4\pm^{0.5}_{0.4} $ & $13.0\pm4.0 $ & $ 4.3\pm1.3 $ \\ 
CE & $ 1.9\pm1.8 $ & $ 1.7\pm^{1.8}_{1.7} $ & $ 3.8\pm3.2 $ & $ 1.3\pm0.8 $ \\ 
DCE & $ 0.0\pm0.0 $ & $ 0.0\pm0.0 $ & $ 1.5\pm1.1 $ & $ 1.5\pm1.1 $ \\ 
Merg & $ 7.0\pm2.6 $ & $ 0.8\pm^{0.8}_{0.8} $ & $54.2\pm2.7 $ & $ 1.9\pm1.3 $ \\ 
\multicolumn{5}{l}{AIC} \\ 
All & $37.4\pm6.3 $ & $34.0\pm4.9 $ & $91.1\pm 11 $ & $42.2\pm7.9 $ \\ 
MT & $ 5.9\pm2.4 $ & $ 5.7\pm2.5 $ & $11.5\pm2.7 $ & $ 7.0\pm2.8 $ \\ 
MT-WD & $28.7\pm2.7 $ & $28.4\pm3.0 $ & $39.9\pm9.5 $ & $31.6\pm9.3 $ \\ 
CE & $ 0.0\pm0.0 $ & $ 0.0\pm0.0 $ & $ 0.8\pm0.4 $ & $ 0.0\pm0.0 $ \\ 
DCE & $ 0.0\pm0.0 $ & $ 0.0\pm0.0 $ & $ 0.6\pm^{0.8}_{0.6} $ & $ 0.6\pm^{0.8}_{0.6} $ \\ 
Merg & $ 2.8\pm1.8 $ & $ 0.0\pm0.0 $ & $38.4\pm5.7 $ & $ 3.0\pm1.2 $ \\ 
\multicolumn{5}{l}{MIC} \\ 
All & $ 1.7\pm1.4 $ & $ 1.1\pm^{1.2}_{1.1} $ & $26.7\pm4.2 $ & $ 3.8\pm2.4 $ \\ 
MT & $ 0.8\pm^{0.8}_{0.8} $ & $ 0.8\pm^{0.8}_{0.8} $ & $ 2.3\pm1.1 $ & $ 0.9\pm^{0.9}_{0.9} $ \\ 
MT-WD & $ 0.4\pm^{0.8}_{0.4} $ & $ 0.4\pm^{0.8}_{0.4} $ & $ 3.4\pm2.5 $ & $ 1.1\pm0.8 $ \\ 
CE & $ 0.0\pm0.0 $ & $ 0.0\pm0.0 $ & $ 1.7\pm1.4 $ & $ 0.2\pm^{0.4}_{0.2} $ \\ 
DCE & $ 0.0\pm0.0 $ & $ 0.0\pm0.0 $ & $ 0.4\pm^{0.5}_{0.4} $ & $ 0.4\pm^{0.5}_{0.4} $ \\ 
Merg & $ 0.6\pm0.5 $ & $ 0.0\pm0.0 $ & $18.9\pm4.4 $ & $ 1.1\pm^{1.2}_{1.1} $ \\ 
\multicolumn{5}{l}{Recently formed or mildly recycled high $B$-field}  \\ 
All & $ 0.2\pm0.0 $ & $ 0.2\pm0.0 $ & $ 2.9\pm0.2 $ & $ 0.6\pm0.1 $ \\ 
MT & $ 0.0\pm{0.0} $ & $ 0.0{0.0} $ & $ 0.1\pm0.0 $ & $ 0.1\pm0.0 $ \\ 
MT-WD & $ 0.2\pm0.0 $ & $ 0.2\pm0.0 $ & $ 0.5\pm0.2 $ & $ 0.4\pm0.1 $ \\ 
CE & $ 0.0\pm0.0 $ & $ 0.0\pm0.0 $ & $ 0.0\pm0.0 $ & $ 0.0\pm0.0 $ \\ 
DCE & $ 0.0\pm0.0 $ & $ 0.0\pm0.0 $ & $ 0.0\pm0.0 $ & $ 0.0\pm0.0 $ \\ 
Merg & $ 0.0\pm{0.0} $ & $ 0.0\pm0.0 $ & $ 2.3\pm0.2 $ & $ 0.1\pm0.0 $ \\ 

\hline
\end{tabular}
\medskip

The number of pulsars that are retained in the halo and the core of our Terzan 5 model.
We indicate pulsars formed via CC, EIC, AIC or MIC, and that gained mass through different
mechanisms (a mass transfer, a common envelope or a merger).
\end{table}

\begin{table}
\caption{Types of pulsars at 11 Gyr, 47 Tuc.}
\label{tab-puls-zoo-tuc}
\begin{tabular}{@{}l c c c c }
\hline
 & \multicolumn{2}{c}{Halo} & \multicolumn{2}{c}{Core} \\
& Total & Binary  & Total & Binary \\
\hline
\\
\multicolumn{5}{l}{All ``plausible''  pulsars} \\ 
All & $ 146\pm 15 $ & $84.8\pm4.7 $ & $ 180\pm 17 $ & $77.1\pm 11 $ \\ 
MT & $37.1\pm7.9 $ & $36.7\pm7.3 $ & $26.5\pm5.0 $ & $18.8\pm5.7 $ \\ 
MT-WD & $40.8\pm5.9 $ & $40.8\pm5.9 $ & $51.8\pm 16 $ & $39.6\pm 14 $ \\ 
CE & $ 8.6\pm5.3 $ & $ 7.3\pm4.2 $ & $ 7.7\pm2.7 $ & $ 3.7\pm2.2 $ \\ 
DCE & $ 0.0\pm0.0 $ & $ 0.0\pm0.0 $ & $ 8.2\pm2.9 $ & $ 8.2\pm2.9 $ \\ 
Merg & $59.5\pm 11 $ & $ 0.0\pm0.0 $ & $86.1\pm6.1 $ & $ 6.9\pm3.1 $ \\ 
\multicolumn{5}{l}{Core Collapse} \\ 
All & $22.4\pm8.8 $ & $ 1.6\pm^{2.7}_{1.6} $ & $ 6.9\pm6.0 $ & $ 2.0\pm2.0 $ \\ 
MT & $ 1.2\pm^{1.8}_{1.2} $ & $ 1.2\pm^{1.8}_{1.2} $ & $ 0.4\pm^{0.9}_{0.4} $ & $ 0.0\pm0.0 $ \\ 
MT-WD & $ 0.4\pm^{0.9}_{0.4} $ & $ 0.4\pm^{0.9}_{0.4} $ & $ 0.4\pm^{0.9}_{0.4} $ & $ 0.4\pm^{0.9}_{0.4} $ \\ 
CE & $ 0.8\pm^{1.1}_{0.8} $ & $ 0.0\pm0.0 $ & $ 0.0\pm0.0 $ & $ 0.0\pm0.0 $ \\ 
DCE & $ 0.0\pm0.0 $ & $ 0.0\pm0.0 $ & $ 0.8\pm^{1.1}_{0.8} $ & $ 0.8\pm^{1.1}_{0.8} $ \\ 
Merg & $20.0\pm6.0 $ & $ 0.0\pm0.0 $ & $ 5.3\pm^{5.5}_{5.3} $ & $ 0.8\pm^{1.1}_{0.8} $ \\ 
\multicolumn{5}{l}{ECS} \\ 
All & $33.4\pm6.2 $ & $ 8.2\pm4.3 $ & $77.9\pm 14 $ & $22.0\pm 11 $ \\ 
MT & $ 1.2\pm^{1.8}_{1.2} $ & $ 0.8\pm^{1.1}_{0.8} $ & $ 8.2\pm2.5 $ & $ 5.3\pm2.3 $ \\ 
MT-WD & $ 0.0\pm0.0 $ & $ 0.0\pm0.0 $ & $ 7.4\pm3.1 $ & $ 4.5\pm3.9 $ \\ 
CE & $ 7.7\pm4.6 $ & $ 7.3\pm4.2 $ & $ 6.5\pm2.7 $ & $ 3.3\pm2.7 $ \\ 
DCE & $ 0.0\pm0.0 $ & $ 0.0\pm0.0 $ & $ 5.7\pm1.7 $ & $ 5.7\pm1.7 $ \\ 
Merg & $24.5\pm4.8 $ & $ 0.0\pm0.0 $ & $50.2\pm9.6 $ & $ 3.3\pm2.3 $ \\ 
\multicolumn{5}{l}{AIC} \\ 
All & $90.1\pm3.6 $ & $75.0\pm6.4 $ & $88.5\pm 23 $ & $51.0\pm 15 $ \\ 
MT & $34.7\pm5.0 $ & $34.7\pm5.0 $ & $17.9\pm6.7 $ & $13.5\pm6.0 $ \\ 
MT-WD & $40.4\pm6.5 $ & $40.4\pm6.5 $ & $43.2\pm 15 $ & $34.3\pm 15 $ \\ 
CE & $ 0.0\pm0.0 $ & $ 0.0\pm0.0 $ & $ 1.2\pm1.1 $ & $ 0.4\pm^{0.9}_{0.4} $ \\ 
DCE & $ 0.0\pm0.0 $ & $ 0.0\pm0.0 $ & $ 0.4\pm^{0.9}_{0.4} $ & $ 0.4\pm^{0.9}_{0.4} $ \\ 
Merg & $15.1\pm4.7 $ & $ 0.0\pm0.0 $ & $25.7\pm 11 $ & $ 2.5\pm1.7 $ \\ 
\multicolumn{5}{l}{MIC} \\ 
All & $ 0.0\pm0.0 $ & $ 0.0\pm0.0 $ & $ 6.9\pm2.3 $ & $ 2.0\pm^{2.0}_{2.0} $ \\ 
MT & $ 0.0\pm0.0 $ & $ 0.0\pm0.0 $ & $ 0.0\pm0.0 $ & $ 0.0\pm0.0 $ \\ 
MT-WD & $ 0.0\pm0.0 $ & $ 0.0\pm0.0 $ & $ 0.8\pm^{1.8}_{0.8} $ & $ 0.4\pm^{0.9}_{0.4} $ \\ 
CE & $ 0.0\pm0.0 $ & $ 0.0\pm0.0 $ & $ 0.0\pm0.0 $ & $ 0.0\pm0.0 $ \\ 
DCE & $ 0.0\pm0.0 $ & $ 0.0\pm0.0 $ & $ 1.2\pm^{1.8}_{1.2} $ & $ 1.2\pm^{1.8}_{1.2} $ \\ 
Merg & $ 0.0\pm0.0 $ & $ 0.0\pm0.0 $ & $ 4.9\pm1.1 $ & $ 0.4\pm^{0.9}_{0.4} $ \\ 
\multicolumn{5}{l}{Recently formed or mildly recycled high $B$-field}  \\ 
All & $ 0.1\pm0.1 $ & $ 0.1\pm0.1 $ & $ 1.8\pm0.3 $ & $ 0.8\pm0.1 $ \\ 
MT & $ 0.1\pm0.1 $ & $ 0.1\pm0.1 $ & $ 0.2\pm0.2 $ & $ 0.2\pm^{0.2}_{0.2} $ \\ 
MT-WD & $ 0.0\pm{0.0} $ & $ 0.0\pm{0.0} $ & $ 0.5\pm0.2 $ & $ 0.4\pm0.2 $ \\ 
CE & $ 0.0\pm0.0 $ & $ 0.0\pm0.0 $ & $ 0.0\pm0.0 $ & $ 0.0\pm0.0 $ \\ 
DCE & $ 0.0\pm0.0 $ & $ 0.0\pm0.0 $ & $ 0.0\pm0.0 $ & $ 0.0\pm0.0 $ \\ 
Merg & $ 0.0\pm{0.0} $ & $ 0.0\pm0.0 $ & $ 1.0\pm0.1 $ & $ 0.1\pm0.1 $ \\ 

\hline
\end{tabular}
\medskip

Notations are as in Table~\ref{tab-puls-zoo-ter}, but for our model of 47 Tuc.

\end{table}

{
\begin{figure}
\includegraphics[height=.28\textheight]{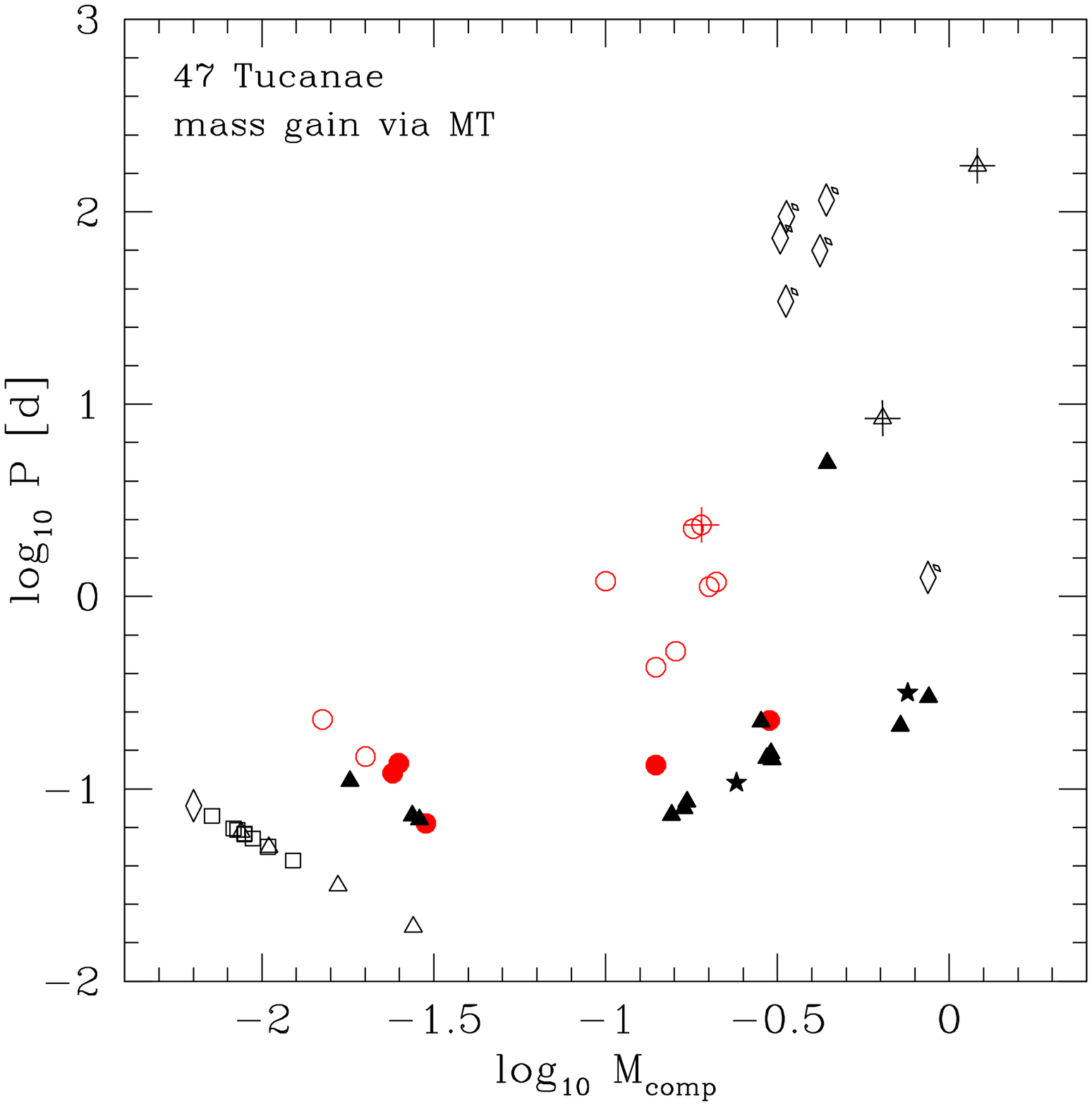} \\
\includegraphics[height=.28\textheight]{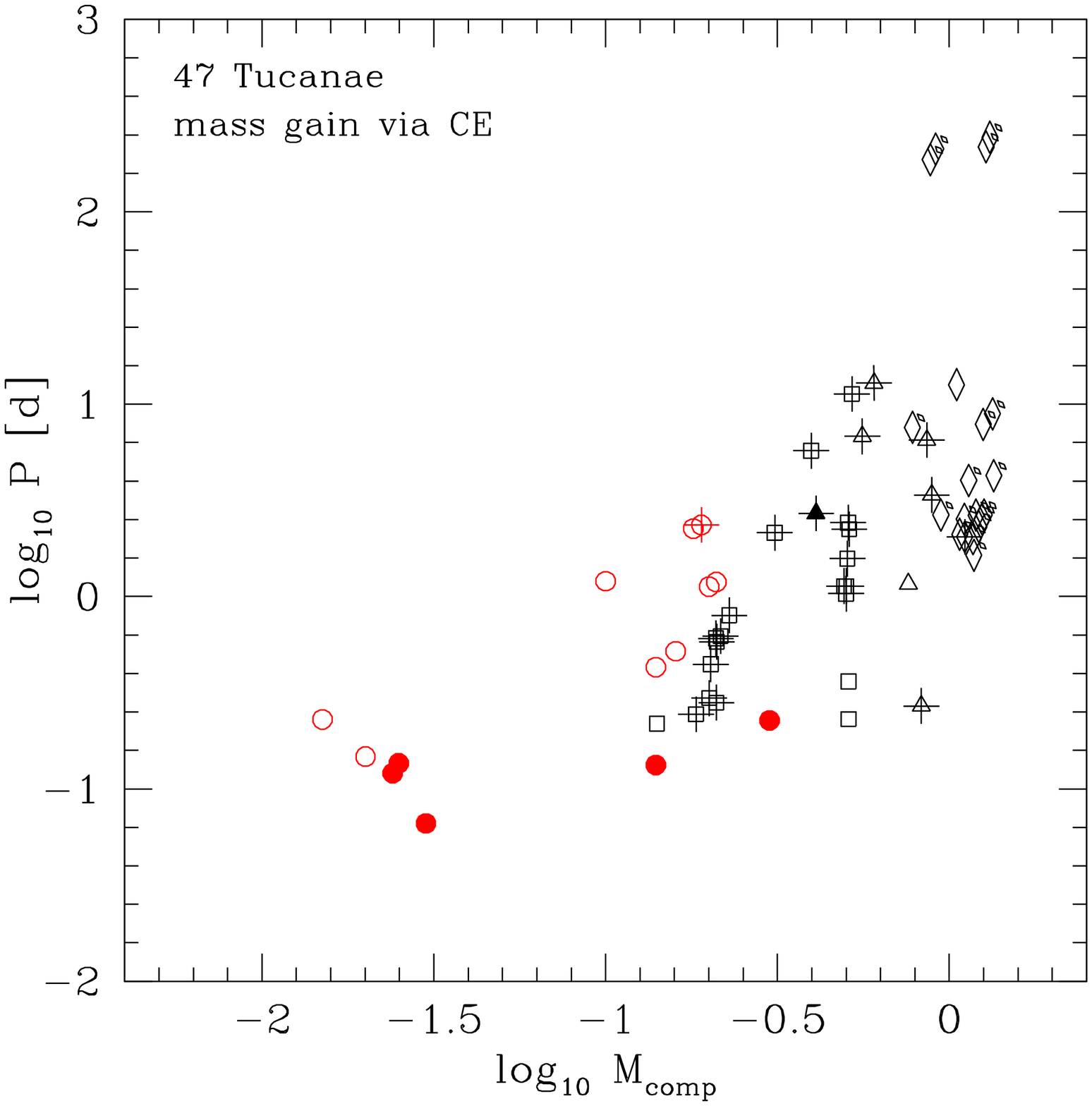} \\
\includegraphics[height=.28\textheight]{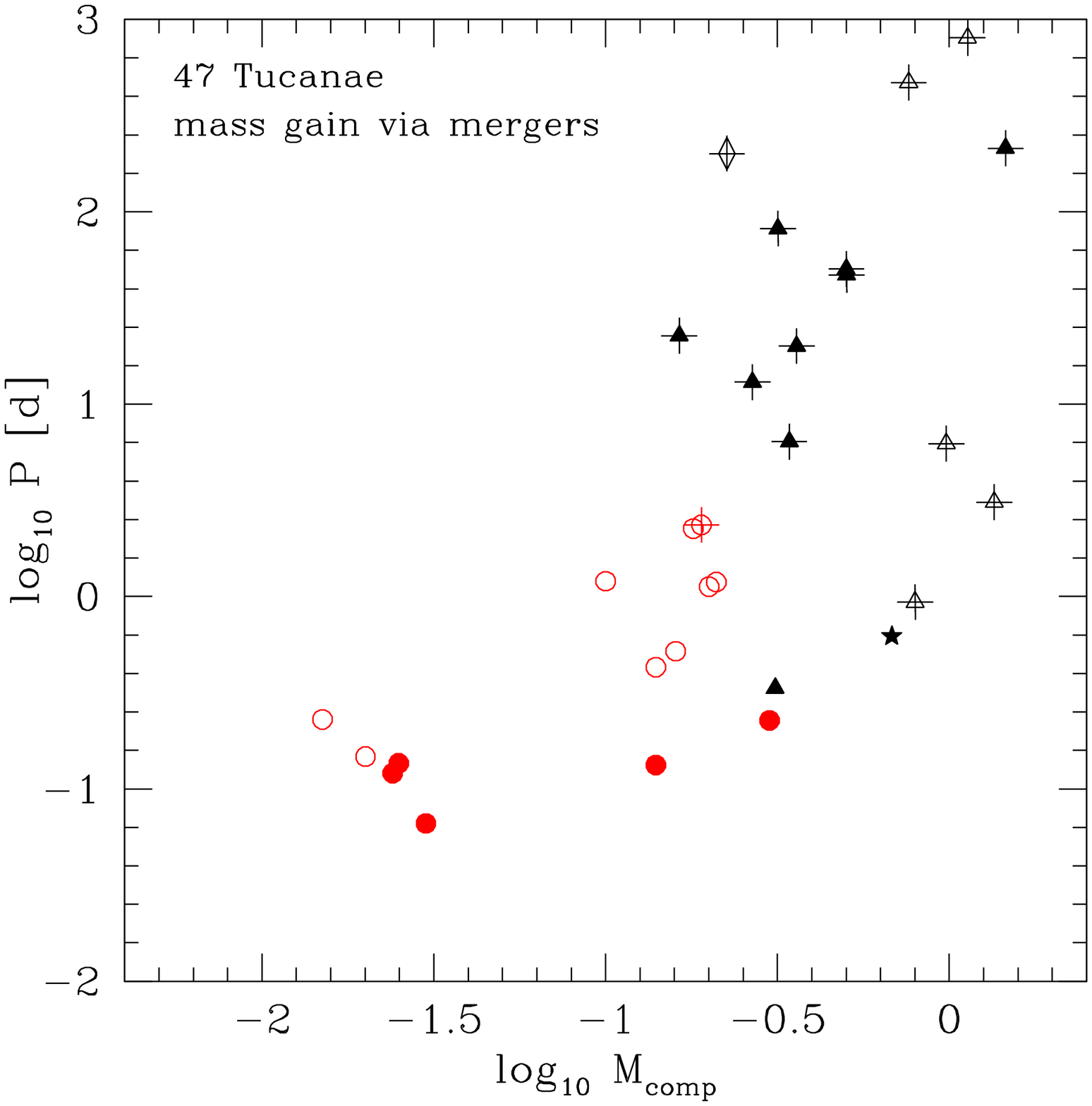} \\
\caption{
Binary MSPs in 47 Tucanae. Observed and modeled populations: circles -- observed, stars -- formed via TC,
squares -- formed via DCE, triangles -- formed via binary encounters and diamonds -- ``primordial'' binary MSPs.
Filled symbols represent binary MSPs with non-WD companions; in the case of observed systems filled symbols
represent eclipsing bMSPs.
Crosses mark eccentric bMSPs ($e>0.05$), adjacent small rotated diamonds mark bMSPs in the halo. 
The modeled population represents accumulated results from 5 models and corresponds to a stellar population twice as large as 47~Tucanae.
\label{fig-msp-tuc}
}
\end{figure}
\begin{figure}
\includegraphics[height=.28\textheight]{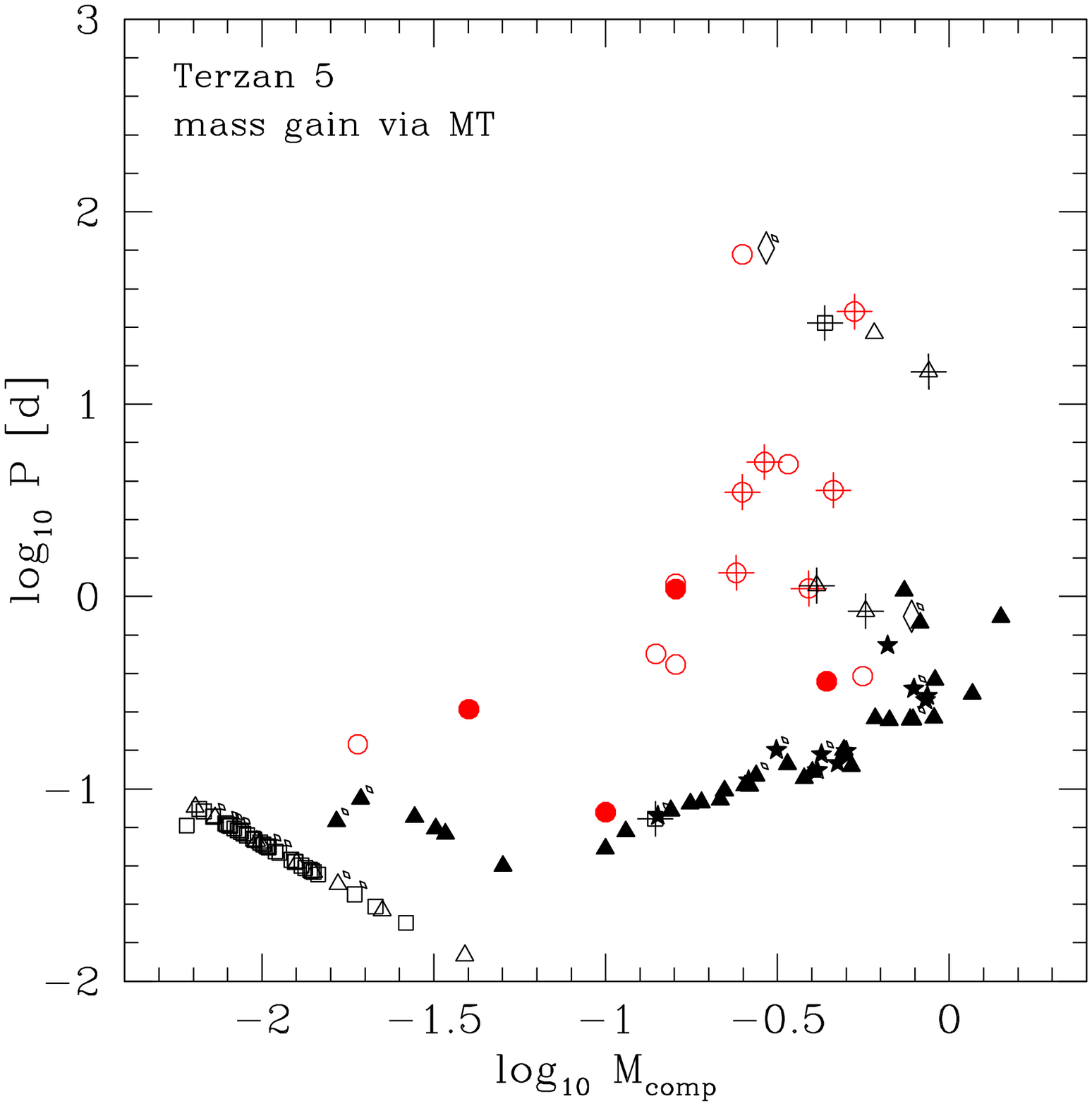} \\
\includegraphics[height=.28\textheight]{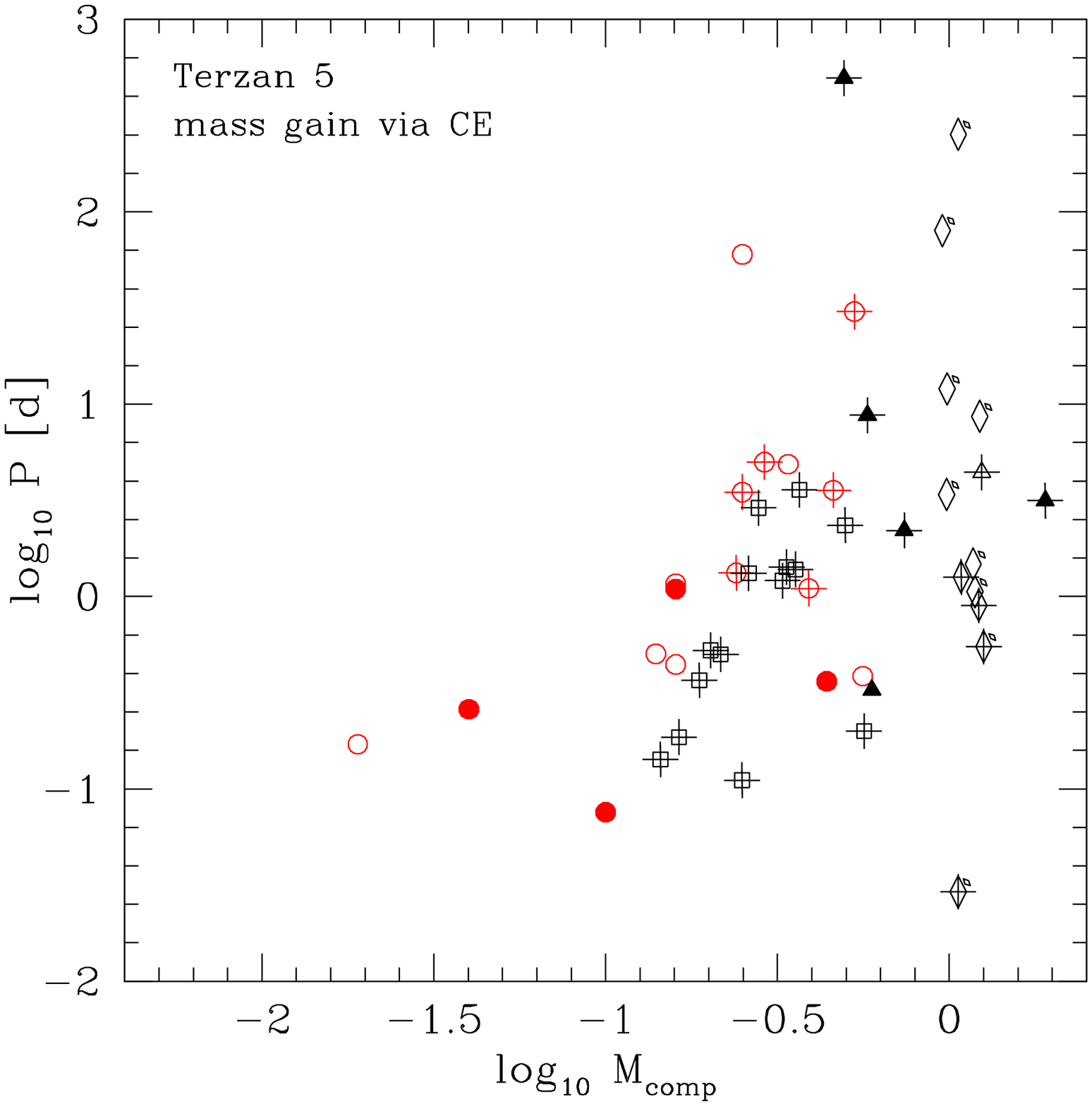} \\
\includegraphics[height=.28\textheight]{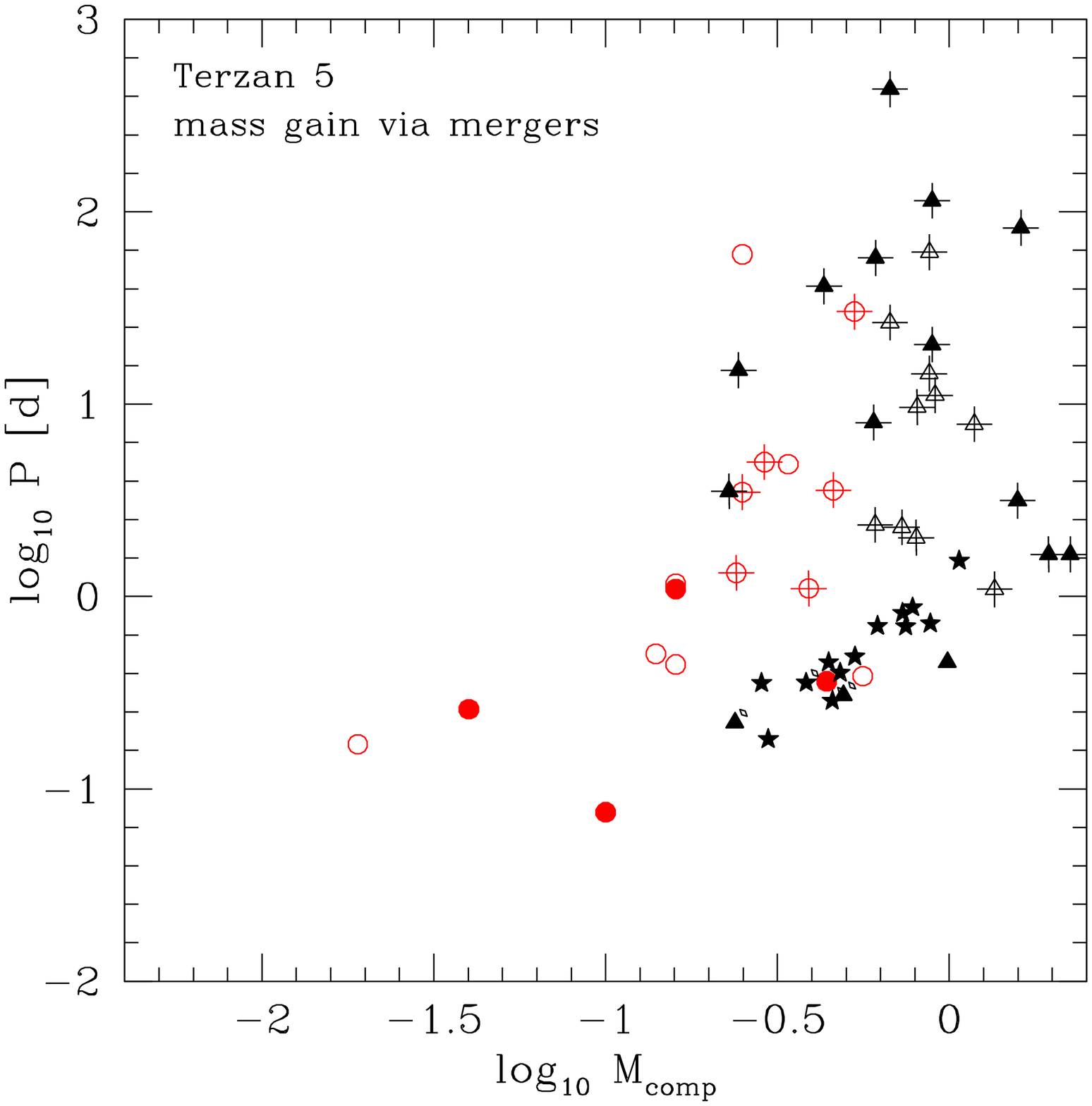} \\
\caption{
Binary MSPs in Terzan~5. Observed (circles) and modeled  populations, symbols as in Fig.~\ref{fig-msp-tuc}.
The modeled population represents accumulated results from 5 models and corresponds to a stellar population 5 times bigger than Terzan~5.
\label{fig-msp-ter}
}
\end{figure}
}

In Figs.~\ref{fig-msp-tuc} and \ref{fig-msp-ter} we compare the simulated populations of NSs that have gained mass 
with the observations of 47 Tuc and Terzan 5.
We use all simulated models for those clusters, and, as a result, the  
simulated populations plotted represent stellar populations about 2 times larger than  47~Tuc and 
5 times larger than Terzan~5.
There are three panels for each cluster. Each panel shows the observed population, compared to  
the simulated population of pulsars that were spun up 
 via a different mechanism:
(i) stable MT; (ii) CE or dynamical common envelope (DCE) during a physical collision with a giant; and (iii) as a result of a merger. 
For most observed MSPs, the companion masses are calculated assuming a pulsar mass of 1.35 solar masses 
and an inclination of 60 degrees\footnote{See http://www2.naic.edu/~pfreire/GCpsr.html}, 
leading to some uncertainty in the direct comparison between observations and simulations.  
In the following we consider MSP populations, grouped by binary formation mechanism. 

\subsubsection{Primordial bMSPs}

This population is present on the panels that show pulsars that gain mass
via MT and via CE. The first population (MT) is characterized by rather large periods, small companion masses, and low eccentricities.
 Systems which may have been formed in this way exist in Terzan 5 (Ter 5 E), M53 (B1310+18), and M4 (B1620-26; 
note that a dynamical exchange is necessary to add the planet, \cite{Ford00_sec}).

The second population, characterized by a companion mass around
$1 M_\odot$ (see the second panels for each cluster), is however undetected in both Ter 5 and 47 Tuc.  
No promising candidates are found among the entire GC bMSP population.  
Typical progenitors would be initially wide binaries containing two stars of rather similar masses from 6 to 10 $M_\odot$.
Most of the primordial bMSPs are located in the halo. There are several possible explanations
for these bMSPs to not be present in real clusters:

\begin{itemize}
\item accretion during the CE either does not occur or does not lead to NS spinup and MSP formation;
\item as NSs in those bMSPs are mainly post-EIC  (see  Tables~\ref{tab-puls-zoo-ter} and \ref{tab-puls-zoo-tuc}) -- 
perhaps the NS natal kicks are higher, causing such systems to be destroyed rather than evolve via CE;
\item the CE prescription adopted in our code ($\alpha_{\rm CE}\lambda=1$) does not produce realistic results for these systems;
\item our mass segregation does not work properly for these systems -- if the progenitors segregate
into the cluster core faster, then they are destroyed and do not produce bMSPs.
\end{itemize}

\noindent If indeed the kicks for EIC are higher than adopted in our code, then we will be able to 
retain very few NSs, as their main supply is from EIC. This will have a rather dramatic effect on
the formation rate of LMXBs, decreasing it strongly. 
If the CE efficiency is the problem, then its reduction might help, as binaries will be formed with smaller binary separations, and 
will shrink their orbits faster and start MT. 
In our "CE-reduced'' GC model (with CE efficiency is
10 times smaller) their number  is a few times less, \
but still large compared to observations. 
Perhaps CE in these systems must be described by a much smaller $\lambda$ parameter, e.g., \cite{Podsi03_bh} showed 
that $\lambda$ can be as low as 0.01. 
In this case all systems evolved through a CE start MT or merge shortly 
after the CE. 
Alternatively, post-CE systems could be described by the angular momentum loss prescription proposed in \cite{Nelemans05_ce}.
Then post-CE systems will be much wider and thus more easily destroyed through dynamical encounters. 
Whatever the true reason, from now on we assume that in our simulations 
a CE event in primordial binaries with a NS formed via EIC does not lead to the formation of a MSP.

Another population of systems which we do not show in Figs. 6 and 7 are NSs produced in AIC, 
and then mildly recycled in the same binary system (see \S~\ref{msp-mild-rec}). 
An example of a slow, likely post-AIC pulsar, is PSR B1718-19 in NGC 6342 \citep{Lyne93}, with a spin period of 1 s.
Its location well outside the cluster half-mass radius suggests that it may be 
a halo object and primordial pulsar, perhaps produced through AIC. 
It is the only slow globular cluster pulsar with a companion in a circular orbit.  
Its companion's wind eclipses the pulsar, and the companion may be filling its Roche lobe 
\citep{Janssen05}, which would support a recent AIC scenario. 
The cluster NGC 6342 is the least dense cluster inhabited by a slow pulsar, and 
PSR B1718-19 has the shortest life-time ($\sim10^7$ yrs) and highest $B$-field of any GC pulsar.  
Both characteristics suggest that this system may have been produced from a 
different mechanism than the other slow pulsars.  
The short life-time also implies a fairly high formation rate (see Tables 9 and 10).

\subsubsection{ {Dynamical common envelope events}} 

DCE leads at first to the formation of an eccentric binary MSP with orbital periods of 0.1-10 days,
where a NS is spun up during the DCE.   Later this system may start a new MT and become an UCXB.
As a result, the population of DCE bMSPs can be found on two panels in Figs.~\ref{fig-msp-tuc} and \ref{fig-msp-ter} 
-- among the NSs that gained mass via MT and via DCE. 

Binary MSP systems produced in NS-WD systems (from UCXBs) are produced in 
large numbers by our simulations, but no such MSPs have been observed in 
any globular cluster (see Figs.~\ref{fig-msp-tuc} and \ref{fig-msp-ter}).
The possibilities for the fate of these systems are summarized in \S\ref{mt-bmsp}. 
We note here that failure of pulsar detection algorithms to detect their short periods seems an unlikely explanation, since the orbital periods of the majority of our simulated MSPs 
from UCXBs are similar to those of 47 Tuc R and Ter 5 A (see  Figs.~\ref{fig-msp-tuc} and \ref{fig-msp-ter}). 
It seems that UCXBs do not produce observable MSPs in ultracompact binaries.  
In Table~\ref{tab-puls-observ} we retain statistics for these systems (as low-$B$, binary WDs with P<0.1d), 
although we do not know how they truly manifest themselves.

The population of simulated bMSPs produced via DCE nicely explains observations 
of eccentric bMSPs, such as Ter 5 I, J, Q, X, Y, and Z, and 47 Tuc H. 
 Our DCE simulations produce some binaries with low eccentricity, but cannot 
reach $e<10^{-3}$.  
Thus, they cannot explain the low eccentricities of 47 Tuc E, Q, S, T, and U, as well as systems like NGC 6752 A.
A stable mass-transfer mechanism must be invoked here (see discussion below in \S~8.2.4).

\subsubsection{ Tidal captures.} 

This population is present on all panels, as the MSP's spin-up 
generally occurred before the MSP acquired its current partner 
(``5b'' in Fig.~\ref{fig-msp-scenario}).
The eclipsing bMSPs Terzan 5 P ($P_{\rm orb}=0.36$ d, $M_{\rm c}=0.44 M_{\odot}$) and 47 Tuc W ($P_{\rm orb}=0.13$ d, $M_{\rm c}=0.14 M_{\odot}$) may have entered their current configurations through this route.

\subsubsection{  Binary exchanges.} 

The top panels of Figs.~\ref{fig-msp-tuc} and \ref{fig-msp-ter} include those bMSPs  which were exchanged, and then underwent MT. 
While the companion mass remains above $\sim 0.2 M_\odot$, 
the MT rate in these systems is  high, and most of them will be seen as LMXBs or qLMXBs (see also \S\ref{msp-rec-mt}).  
However, the transition from LMXB to MSP is uncertain, and several bMSPs 
with nondegenerate companions are likely to have been formed through binary 
exchange or tidal capture (47 Tuc W and V; Ter 5 A, P, ad, and perhaps O; NGC 6397 A).  
Higher-mass bMSPs with nondegenerate companions may have their radio 
signals continually obscured by ionized material, leading to ``hidden'' bMSPs 
\citep{Tavani91,Freire05}. 

Those bMSPs which underwent CV-like evolution after their exchanges 
can explain the very low-mass (brown dwarf) companions to bMSPs 
(e.g. 47 Tuc I, J, O, P, and R; Ter 5 ae).
CV-like evolution also can lead to the formation of binaries with longer orbital periods (from one to ten days) and 
low-mass WD companions.
We find that binaries with post-MT periods of 5-10 days are formed relatively frequently, 
but all of them have collision times well below 1 Gyr. After the MT, they experienced
a dynamical encounter that led either to an exchange of companions, or to an eccentricity increase.
As a result, none of them are observed at 11 Gyr with low eccentricities. 
Potentially, we should be able to form 
binaries with a post-MT period of about a day. Such binaries have a collision time
around 10 Gyr and thus should stay preserved in their post-MT shape for almost  the entire 
cluster evolution. 
However, probably due to rather low
number statistics, they are not formed in our simulations for 47 Tuc and Ter 5.
Also, our post-exchange binaries are very eccentric. A better treatment of the MT 
in eccentric binaries, like discussed in \cite{Sepinsky07}, 
may help to create low-eccentric binaries with low-mass WD companions.

The bMSPS which were spun-up via mergers during a previous dynamical
interaction (bottom panels) generally have more massive companions and longer 
periods than the observed populations.  
The few good candidates are Ter 5 Q, NGC 1851 A, M28 D, NGC 6441 A, and possibly M30 B.
Dynamically formed binaries which 
undergo stable MT with a giant, producing non-eccentric long-period systems
are rare in our simulations; these tend to come from the primordial population located outside the core.

\begin{table*}
\caption{``Observable'' pulsars at 11 Gyr.}
\label{tab-puls-observ}
\begin{tabular}{@{}l l l l l l l l l l }
\hline
 model & \multicolumn{8}{c}{low-$B$} & {High-$B$}\\
& All psrs & Single &  \multicolumn{6}{c}{Binary} & All psrs\\
Companion& & & BD & WD & & MS & He WD & heavy WD & recently\\
 $M_{\rm c}$& &  & & & & & $\la0.5 M_\odot$& $\ga0.5 M_\odot$& recycled\\
Period & & & $<0.1$d& $<0.1$d& $>10$d & &  $0.1\div10$d&  $0.1\div10$d \\
\hline
standard & $ 5.2\pm1.7 $& $ 2.1\pm1.3 $& $ 0.3\pm0.3 $& $ 1.0\pm0.9 $& $ 0.8\pm0.5 $& $ 0.5\pm0.5 $& $ 1.1\pm^{1.7}_{1.1} $& $ 0.5\pm0.5 $& $ 0.4\pm0.2 $\\ 
metal-poor & $ 3.9\pm2.0 $& $ 2.3\pm1.2 $& $ 0.2\pm^{0.4}_{0.2} $& $ 0.5\pm0.4 $& $ 0.3\pm^{0.4}_{0.3} $& $ 0.2\pm^{0.4}_{0.2} $& $ 0.8\pm0.6 $& $ 0.2\pm^{0.4}_{0.2} $& $ 0.6\pm0.2 $\\ 
high-den & $15.6\pm2.7 $& $ 9.8\pm3.1 $& $ 0.5\pm^{0.8}_{0.5} $& $ 3.5\pm2.0 $& $ 0.0\pm0.0 $& $ 1.4\pm^{1.8}_{1.4} $& $ 3.0\pm0.8 $& $ 1.2\pm^{1.3}_{1.2} $& $ 2.5\pm0.5 $\\ 
med-den & $ 1.4\pm1.5 $& $ 0.6\pm^{1.0}_{0.6} $& $ 0.0\pm0.0 $& $ 0.0\pm0.0 $& $ 0.3\pm^{0.7}_{0.3} $& $ 0.0\pm0.0 $& $ 0.2\pm^{0.4}_{0.2} $& $ 0.3\pm^{0.4}_{0.3} $& $ 0.1\pm0.0 $\\ 
low-den & $ 0.0\pm0.0 $& $ 0.0\pm0.0 $& $ 0.0\pm0.0 $& $ 0.0\pm0.0 $& $ 0.0\pm0.0 $& $ 0.0\pm0.0 $& $ 0.0\pm0.0 $& $ 0.0\pm0.0 $& $ 0.0\pm0.0 $\\ 
low-$\sigma$ & $ 3.4\pm0.9 $& $ 2.2\pm1.3 $& $ 0.3\pm^{0.5}_{0.3} $& $ 0.3\pm^{0.5}_{0.3} $& $ 0.3\pm^{0.5}_{0.3} $& $ 0.7\pm0.7 $& $ 0.2\pm^{0.4}_{0.2} $& $ 0.0\pm0.0 $& $ 0.2\pm0.1 $\\ 
long-$t_{\rm rh}$ & $ 1.6\pm0.8 $& $ 0.8\pm0.6 $& $ 0.2\pm^{0.4}_{0.2} $& $ 0.3\pm^{0.4}_{0.3} $& $ 0.3\pm^{0.7}_{0.3} $& $ 0.0\pm0.0 $& $ 0.2\pm^{0.4}_{0.2} $& $ 0.2\pm^{0.4}_{0.2} $& $ 0.3\pm0.1 $\\ 
BF05 & $ 3.0\pm1.5 $& $ 0.9\pm1.0 $& $ 0.1\pm^{0.3}_{0.1} $& $ 1.0\pm0.4 $& $ 0.7\pm0.0 $& $ 0.4\pm0.4 $& $ 0.3\pm^{0.7}_{0.3} $& $ 0.3\pm^{0.4}_{0.3} $& $ 0.3\pm0.1 $\\ 
fast-MB & $ 8.0\pm3.5 $& $ 2.9\pm1.6 $& $ 1.1\pm0.9 $& $ 1.5\pm1.2 $& $ 1.1\pm0.9 $& $ 1.5\pm1.1 $& $ 1.6\pm1.0 $& $ 0.3\pm^{0.4}_{0.3} $& $ 0.5\pm0.2 $\\ 
CE-reduced & $ 2.4\pm1.7 $& $ 1.3\pm^{1.9}_{1.3} $& $ 0.2\pm^{0.4}_{0.2} $& $ 1.0\pm0.4 $& $ 0.3\pm^{0.4}_{0.3} $& $ 0.2\pm^{0.4}_{0.2} $& $ 0.5\pm^{0.7}_{0.5} $& $ 0.2\pm^{0.4}_{0.2} $& $ 0.4\pm0.2 $\\ 
oldkicks & $ 1.6\pm1.6 $& $ 0.0\pm0.0 $& $ 0.3\pm^{0.4}_{0.3} $& $ 0.0\pm0.0 $& $ 0.8\pm0.8 $& $ 0.5\pm0.4 $& $ 0.3\pm^{0.7}_{0.3} $& $ 0.2\pm^{0.4}_{0.2} $& $ 0.0\pm0.1 $\\ 
47 Tuc & $25.7\pm3.5 $& $ 9.4\pm3.4 $& $ 2.0\pm1.4 $& $ 6.1\pm3.8 $& $ 2.9\pm2.7 $& $ 1.6\pm1.7 $& $ 7.8\pm2.7 $& $ 2.9\pm^{3.1}_{2.9} $& $ 1.8\pm0.2 $\\ 
Terzan 5 & $30.8\pm6.3 $& $22.7\pm2.6 $& $ 2.3\pm1.1 $& $ 9.1\pm2.0 $& $ 0.9\pm^{1.3}_{0.9} $& $ 2.5\pm1.1 $& $ 2.6\pm2.3 $& $ 0.8\pm0.8 $& $ 3.8\pm0.3 $\\ 

\hline
\end{tabular}
\medskip

The numbers of pulsars  that can be detected in a GC (including MSPs only if they gain mass via MT from a non-degenerate donor, or via DCE).
Recent high-$B$ pulsars are those which were formed or mildly recycled less than $10^8$ years ago.
Mass-transferring NS-MS binaries where a MS star is $\ga 0.2 M_\odot$ are excluded, as they are likely LMXBs/qLMXBs.
``Companion'' indicates the type of the companion; 
``$M_{\rm c}$'' is the companion mass, and ``Period'' is the orbital period.
For all GC models, except 47~Tuc and Terzan~5, the numbers 
are scaled per 200 000 $M_\odot$ stellar population mass at the age of 11 Gyr;
for 47 Tuc the numbers are given for 47 Tuc's mass, taken as $10^6$ $M_\odot$;
for Terzan~5, 370 000 $M_\odot$.  
\end{table*}

\subsection{{Refined populations of MSPs and observations}}

From \S\ref{msp-nonrec} and our comparisons with observations, we 
see that not all events where NSs gain mass lead to detectable radio MSPs. 
These events are likely channels for short-lived high-$B$ MSP production: 

\begin{itemize}
\item{recent NS formation via any ECS channel}
\item{AIC followed by continued accretion in the same system}
\item{mass gain via merger or physical collision}
\end{itemize}

\noindent Radio MSPs are not produced via these channels, or can not be detected yet:  
\begin{itemize}
\item{accretion from a degenerate donor}
\item{mass transfer in an NS-MS systems with current MS mass $\ga 0.2M_\odot$, as they have not yet turned on as radio pulsars}
\item{accretion during a CE event}
\end{itemize}

With these restrictions (see Fig.~1), we find that 
our estimates for the numbers of produced MSPs are  
$31\pm6$ in 47~Tuc and $26\pm4$ MSPs in Terzan~5 (see Table~\ref{tab-puls-observ}).  
These estimates are close to the observations for both 47 Tuc (22 detected MSPs) and Terzan 5 (33 MSPs). 

Our simulations indicate that the fraction of single pulsars is higher in Terzan 5 than in 47 Tuc, as observed.
The origin of isolated MSPs depends on the cluster dynamical properties.
For example, in our standard model, about half of all isolated MSPs 
were produced by an evolutionary merger at the end of the MT from a MS companion (\S\ref{mt-bmsp}),
while the other half lost their companion due to a binary encounter (these MSPs were generally spun up by MT from a giant, leading to long periods).
Most isolated pulsars in 47 Tuc lost their companion due to a binary encounter; half of these involved the removal of a low-mass MS star at the end of its MT sequence. 

Spatially, MSPs are more likely to be located where they were formed -- mostly in the core.
Less than a third were produced in or ejected to the halo.
The fraction of MSPs ejected into the halo increases as  $v_{\rm rec}$ decreases.
Roughly half of all observed GC pulsars are found outside their cluster core radius \citep{CamiloRas05}, 
but the radial distribution of most pulsars is exactly as expected for a population 
that is produced in or around the core \citep{Grindlay02, Heinke05a}, 
with the exception of a few, likely ejected, pulsars in M15 and NGC 6752 \citep[e.g.][]{Colpi02}.

The creation rate of  high-magnetic field (AIC) MSPs, at the age of 11 Gyr, is about 20-40 per Gyr in Terzan 5 and 47 Tuc.
This leads to the probability of forming high-magnetic field MSPs ($B\la 10^{12}$-gauss) with a period $\ge 100$ ms 
of less than a few per cent, while for $B\la 10^{11}$-gauss, it is about 50 per cent (see eq.~\ref{eq-msp}).
E.g., one pulsar in Terzan 5 (J1748-2446J) has P=80 ms \citep{Ransom05}.
The absence of slow-period MSPs in 47 Tucanae suggests that the magnetic field in newly formed AIC NSs is most likely to be at
the higher end, $\ga 10^{12}$-gauss. We note that the results of our simulations are
consistent with a higher probability of a high-$B$ pulsar in Terzan 5 than 
in 47 Tucanae (see Table~\ref{tab-puls-observ}), though the statistics are very poor.

We compare the behavior of pulsar properties with increasing cluster density in our simulations, and in real clusters, in Table \ref{tab-craig}.
We divide the clusters containing MSPs with known properties into 
categories roughly corresponding to the central densities used in the models: 
High-density ($\rho_{\rm c} >5.30$, and core-collapsed with $\rho_{\rm c} >5.20$): 
NGC 1851,  NGC 6397, NGC 6522, NGC 6544, NGC 6624, M15, and M30.
Terzan-5-like ($5.05 < \rho_{\rm c} <5.30$): NGC 6266, Terzan 5, NGC 6440, and NGC 6441.
47Tuc-like ($4.6 < \rho_{\rm c} <5.05$): 47 Tuc,  NGC 6342,  NGC 6626, and NGC 6752.
Standard ($3.8 < \rho_{\rm c} < 4.6$): M5, M4, and NGC 6760.
Medium-density ($2.8 < \rho_{\rm c} < 3.8$): M53, M3, M13, NGC 6539, NGC 6749, and M71.
No pulsars have been found to inhabit low-density  ($2.8 > \rho_{\rm c}$) clusters yet, nor do we produce any in our simulations  (Table~\ref{tab-puls-observ}).
We exclude from consideration all MSPs with unknown binary properties, and those isolated pulsars in clusters which have been observed too rarely to measure binary properties (NGC 6624 D and E; NGC 6517 A and C) to reduce bias towards single MSPs. 

\begin{table*}
\caption{Detected pulsars and simulations.}
\label{tab-craig}
\begin{tabular}{@{}l c c c c c c c c c}
\hline
Cluster      &  All pulsars  & Single  & BD  & He WD  & MS & heavy WD & $>10$ d & ecc & high-B \\
\hline
High-den model & $18.8\pm3.2 $& $0.66\pm0.11 $& $0.03\pm0.03 $& $0.16\pm0.05 $& $0.08\pm^{0.09}_{0.08} $& $0.10\pm0.08 $& $0.01\pm0.00 $& $0.19\pm0.10 $& $0.14\pm0.04 $\\ 
All high-den   &  19       & 0.71 & 0.06       & 0.0 & 0.12 & 0.12 & 0.06 & 0.18 & 0.12 \\
\hline
Terzan 5 model    & $35.7\pm6.4 $& $0.75\pm0.06 $& $0.05\pm0.03 $& $0.08\pm0.06 $& $0.07\pm0.03 $& $0.07\pm0.03 $& $0.03\pm0.03 $& $0.12\pm0.07 $& $0.11\pm0.02 $\\
All Ter5    & 49           & 0.43 & 0.08    & 0.31 & 0.12 & 0.06 & 0.08 & 0.16 & 0.04 \\
\hline
47Tuc model & $28.3\pm3.1 $& $0.38\pm0.06 $& $0.07\pm0.04 $& $0.35\pm0.09 $& $0.07\pm0.06 $& $0.14\pm0.12 $& $0.10\pm0.10 $& $0.37\pm0.14 $& $0.06\pm0.01 $ \\
All 47Tuc   & 37           & 0.43 & 0.19    & 0.30 & 0.11 & 0 & 0.03 & 0.08 & 0.03 \\
\hline
Standard model & $ 5.8\pm1.7 $& $0.44\pm0.18 $& $0.04\pm0.04 $& $0.27\pm0.17 $& $0.12\pm0.10 $& $0.15\pm0.10 $& $0.16\pm0.10 $& $0.35\pm0.20 $& $0.07\pm0.03 $\\
all stand  & 8            & 0.25 & 0.25    & 0.50 & 0.0 & 0 & 0.25 & 0.25 & 0.0 \\
\hline
Med-den model &   $ 1.5\pm1.5 $& $0.23\pm^{0.28}_{0.23} $& $0.02\pm^{0.04}_{0.02} $& $0.27\pm^{0.39}_{0.27} $& $0.02\pm^{0.04}_{0.02} $& $0.28\pm^{0.41}_{0.28} $
& $0.09\pm^{0.18}_{0.09} $& $0.22\pm^{0.39}_{0.22} $& $0.07\pm0.03 $\\
All med-den   & 11          & 0.18 & 0.18    & 0.64 & 0 & 0 & 0.18 & 0.18  & 0 \\
\hline
Field        & 77          & 0.25 & 0.04    & 0.53 & 0 & 0.13 & 0.30 & 0.01 & - \\
\hline
\end{tabular}
\medskip
Fractions of GC pulsars by system type, simulations vs. observations for each 
central density.  Observations summarized by \citet{CamiloRas05}, 
with additions from http://www.naic.edu/~pfreire/GCpsr.html as of 
late 2007.  See text (\S 8) for details.  

\end{table*}

The column ``All pulsars'' indicates the total number of pulsars used in that 
category; other columns refer to fractions of that number.  The total numbers 
of pulsars cannot be usefully compared to the numbers of pulsars in the model, 
but the fractions of pulsar system types can be compared. 
Field systems include all field pulsars with $P<100$ ms and $B<5\times10^{10}$ G 
\citep{Manchester05}\footnote{http://www.atnf.csiro.au/research/pulsar/psrcat}. 
``BD'' refers to brown dwarf systems; $M_{\rm comp}<0.07\ M_\odot$.  ``MS'' identifies likely main-sequence companions, identified by being eclipsing systems with  $M_{\rm comp}>0.07\ M_\odot$ \citep{Freire05}. 
``He WDs'' refer to all non-eclipsing systems with $0.07<M_c<0.5\ M_\odot$.
``Heavy WDs'' include all MSPs with $M_c\ga 0.5\ M_{\odot}$, unless the companion 
is known to be a NS.  ``$>10$ d'' refers to that subset of all MSPs 
with orbital periods longer than 10 days.  
``High-$B$'' pulsars are taken to be MSPs with spin-period $P\ga 100$ ms. 
To improve the statistics for our high-$B$ systems we count  
all high-$B$ pulsars  spun up within the past 1 Gyr, and divide  
this number by a factor of 10 to account for their shorter lifetimes.

 Several clear trends with the central density of globular clusters 
can be seen in the observations.  Single MSPs rise 
smoothly from a rate of 0.25 of all MSPs in the field, to 0.7 of all 
MSPs in clusters.  
This is well-described by our model, suggesting that the formation of single MSPs may be reasonably well understood.
The formation of MSPs with ``brown dwarf'' companions, with very low masses, 
exhibit a complicated density dependence, with few in the field, a quarter of all MSPs in standard-density 
clusters, and declining fractions in higher density clusters.  
This indicates that their formation requires a high density environment, 
but also suggests that their binary evolution can be disrupted in high 
density conditions.  
Our simulations do not reproduce this trend very well; likely this 
is because our lack of a proper equation of state for brown dwarfs 
prevents us from accurately tracking the late evolution of NS-MS binaries 
and the eventual possible destruction of the companion.  
We may produce too many heavy WDs and too few He WDs, which both have only mild density dependences.   
The fraction of eccentric MSPs, perhaps surprisingly, shows little density dependence (in our simulations and in real systems). 
The production of MSPs with main-sequence companions 
and of high-$B$ MSPs are sharply dependent 
on the central density of the cluster, in agreement with the suggestions
by \citet{Freire05} and \citet{Lyne96} 
that their production requires multiple binary interactions, and 
direct impact of NSs with other stars, respectively.
Overall, our predicted numbers are in quite good agreement with the 
observations.

\section{Double neutron stars in clusters}

As predicted in \S\ref{ns_prim}, we did not find primordial double neutron star formation to be efficient.
In all our simulations -- 70 cluster models or $2\times 10^7 M_\odot$ of stellar population evolved in different clusters --
only 3 primordial DNSs were formed. In all three systems both NSs were formed via ECS, and none of the systems have merged
within a Hubble time.

Let us estimate the rate of dynamical DNS formation. 
The characteristic time for a single NS to have an encounter with a NS-binary 
can be estimated  using  eq.~\ref{eq-tau-coll}. 
The binary companion in a NS-binary is usually less massive than the NS.
If an exchange occurs, the binary separation in the post-exchange binary $a$ 
is larger than the binary separation $a_0$ in the pre-exchange binary,
$a\sim a_{0} m_{\rm NS}/m_{2}$, where $m_2$ is the mass
of the replaced companion \citep{Heggie96}.
Binary eccentricities in hard binaries follow a thermal distribution,
with an average eccentricity  $e\sim 2/3$ \citep{Heggie75}.
The maximum $a$ that leads to a merger within 11 Gyr, 
if the post-exchange eccentricity is 2/3, is about $7.5 R_\odot$. 
The maximum mass for a NS-companion before the exchange is about the mass of a turn-off MS star -- $0.9 M_\odot$.
Then the maximum binary separation in a pre-exchange binary that leads to the formation of a merging DNS is $a_0=4.8 R_\odot$. 

From our simulations we find that the relative fraction of NS binaries in the core 
$f_{\rm NS-bin, core}=N_{\rm NS-bin, core}/N_{\rm objects, core}$ is 0.175 per cent 
for the standard model, and 0.6 per cent for our Terzan 5 model. 
Then the characteristic time for a NS to have an encounter that may lead to the formation of a merging DNS
is $\tau_{\rm DNS,m}\approx5\times10^{12}$ Gyr  in the case of a standard cluster,
in other words, about 500 NSs are needed to form one merging DNS within 11 Gyr.
This assumed that {\it every} encounter will lead to the exchange and 
that  {\it all} NS-binaries have the separation of $\sim 5 R_\odot$. 
In the case of Terzan 5, the chance of forming a merging DNS, per NS, is 25 times bigger than in a standard cluster --
the formation rate is highly dependent on the cluster density, as the power 2, and the number of retained
NSs increases as the escape velocity increases.
However, most NS-binaries have larger binary separations than $\sim 5 R_\odot$,
and even though encounters are more frequent in this case, they do not lead to the formation of a merging DNS directly. 
Subsequent encounters of the wide DNS with other stars would be necessary to increase the DNS's eccentricity, and thus reduce its merging time.
  
Only 14 DNSs are formed dynamically during 11 Gyr in all of our 70 cluster models.
Since the formation rate of DNSs is highly density dependent, 
most of those DNSs were formed in the cluster models of Terzan 5.
Only two of these 14 DNSs merged within 11 Gyr, while another two DNSs merged within 14 Gyr, a Hubble time.
5 of the 14 DNSs stayed in the cluster for 11 Gyr, while the remaining 9 were destroyed within a Gyr after their formation.

We note that the formation rate of DNSs is strongly connected to the formation rate of LMXBs,
as the main building material of DNSs is NS-binaries that otherwise become LMXBs.
As we show in \S\ref{sec-lmxbs}, we produce LMXBs in numbers comparable to those observed,
but we underestimate the formation rates for core-collapsed clusters.
In core-collapsed clusters it is especially clear that our simplified models do not properly treat the
cluster dynamics
and therefore we cannot be very quantitative about the total number of DNSs that should be produced
by all GCs, although only a small fraction of GCs are core-collapsed.  
We conclude therefore that GCs are rather unimportant for the formation of
merging DNSs and can at most provide a small ($10-30\%$) contribution to
the production of short $\gamma$-ray bursts in old host galaxies as
estimated by \cite{Grindlay06}. 
Such systems
can only be produced in core-collapsed clusters, like M15, where such a system is observed \citep{Anderon90_dns,Jacoby06_dns}.
Also, recent observations seem to indicate that
majority of short GRBs are associated with young hosts \citep{Berger07}.
If in fact short hard GRBs originate from NS-NS mergers \citep[although see][]{Belczynski07_nsns}, 
then their rates and occurrence in young hosts is
understood in terms of NS-NS formation without dynamical interactions, i.e.
in field populations (Belczynski, Stanek \& Fryer 2007).

\section{Summary}

In our study of binaries with neutron stars in globular clusters we 
considered in detail the problem of the formation of LMXBs and MSPs. 
We predict from our simulations that most retained neutron stars in GCs
must come from an electron capture supernova formation channel (c.f. the field,
where most NSs are from core-collapse supernovae). 
A typical GC could contain at present (where the adopted cluster age is 11 Gyr) 
as many as $\sim 220$ NSs, of which half are located in the halo; 
a massive GC like 47 Tuc could have more than a thousand NSs.

Analyzing encounters with NSs, we find that for NSs located in the core,
about half formed a binary through an exchange encounter and only a few per cent formed a binary through a physical collision with a giant or via tidal capture. The relative importance of tidal captures 
to physical collisions increases as the velocity dispersion decreases.
Although there are fewer binaries formed through DCE and TC than through binary encounters, the production of mass-transferring systems is roughly the same for all three channels.  
We note as well that many physical collisions lead to mergers between NSs and other stars in the core. 

We derived a ``collision number'' from our globular cluster models which appears to scale linearly with the LMXB production frequency from our simulations, 
although only when the core density alone is varied.
Variations of other GC dynamical properties with a fixed core density
lead to a large scatter. 
This may explain the observed deviations from a linear dependence between the collision number and the number of LMXBs in non-core-collapsed clusters. 
We predict that the numbers of qLMXBs will have no dependence on metallicity, 
in contrast to the numbers of bright LMXBs.

Our rates of LMXB formation predict 7.5 UCXBs in all galactic GCs, which is consistent 
with the observed number. For qLMXBs, we expect about 180 systems, which 
agrees with the range (100-200) estimated from observations \citep{Pooley03,Heinke03a,Heinke05b}. 
From our list of clusters which are expected to have one or more LMXBs (bright or in quiescence)
-- 47 Tuc, Terzan 5, NGC 1851, NGC 6266, NGC 6388, NGC 6440, NGC 6441, NGC 6517, M54, NGC 7078 --
all clusters which have been studied with Chandra (all but NGC 6517) show at least one LMXB or qLMXB.
Our results do not support the idea that the observed luminosity function in extragalactic globular clusters is 
principally due to UCXBs.

The resulting retention fraction of  $\sim 5-7\%$ (or $\sim 220$ NSs per 200,000 $M_\odot$), obtained in our simulations, 
seems to be firm, as the numerically derived formation rates of LMXBs, which strongly correlate 
with the retention fraction, are consistent with those observed.
In our GC models, we achieved such a retention fraction by producing NSs via different channels of electron capture supernova.
If the natal kick velocity distribution  will be revised in future from that used in this paper \citep{Hobbs05_kicks}, then the necessity of ECS will decrease.
If, however, it should be shown that AIC does not occur, then the retention fraction that is required to match the observations of LMXBs might have to be larger by a factor of 2.

We find from our simulations that if all possible channels of NS formation and all possible mechanisms for their spin-up
lead to MSP formation, then we overproduce MSPs.   
However, we still need these channels to produce observed LMXBs.  
We propose that high $B$-field MSPs (which are short-living) can be formed not only during core-collapse supernovae, but also due to physical collisions or accretion in a post-AIC system. 
We find that NSs which accrete and spin up during CE events overproduce 
bMSPs in the cluster halos from primordial binaries of intermediate masses.
Such bMSPs would be present in low-density clusters and have not yet been seen.
In the case of the NS-WD LMXBs, we propose that the MT in such systems does not lead to the formation of radio bMSPs. 
The rates of UCXB formation (verified by observations of UCXB LMXBs) predict 
large numbers of ultracompact bMSPs in GCs which are not detected.  

Excluding the systems discussed above, as well as those which are still actively accreting their donor's material 
and are seen instead as LMXBs, we obtain lower limits 
 of ``detectable'' bMSPs.
The predicted numbers are in rather good agreement with the observations --
$26\pm4$ MSPs in 47 Tuc and $31\pm6$ in Terzan 5.
The fraction of isolated pulsars is comparable to that observed and is larger in Terzan 5 than in 47 Tuc.  

Comparing the population census of our models with the observations of all detected pulsars to date in GCs,
we find good agreement for all types of pulsars -- single, those with brown dwarf companions, with MS companions,
with light or heavy WD companions, and with large periods.

We do not find very efficient formation rates for double NSs -- only a dozen were formed in 70 simulated models.
These rates increase with the square of the GC core density.
We conclude that DNS formation is most likely to occur in massive, very dense,
and preferentially core-collapsed GCs, as suggested by the identification of only one double NS so far, in the core-collapsed GC M15.

In conclusion we outline several important issues that must be addressed
for further progress in studies of NSs in globular clusters: 

\begin{itemize}

\item Common envelope events will occur in primordial binaries of intermediate mass that produce NSs via EIC.  What is the common envelope efficiency, and is the NS spun up by accreting the material?

\item What is the result of mass accretion onto a NS after it has experienced 
a merger, either in a binary due to unstable mass transfer, 
or during a collision? 

\item What is the final fate of a mass-transferring NS-WD binary? 
 
\item How does the evolution of a  mass-transferring NS-MS binary proceed when the companion's mass reaches $\la 0.04 M_\odot$?  
 
\item What is the dependence of qLMXB numbers on the cluster metallicity?  This holds the potential to determine the cause of the LMXB metallicity dependence (see \S~7).

 In particular, if future observations show that the metallicity dependence of qLMXBs follows 
that for bright LMXBs, it will most likely indicate that these two types of clusters had 
different initial mass functions (and thus star formation histories), 
as predicted by \citet{Grindlay93} and expected from star formation theory 
in dense metal-rich environments \citep{Murray07}.
If metal-poor GCs  have similar numbers of MSPs as metal-rich GCs, but fewer qLMXBs 
(indicating a shorter LMXB lifetime), then the idea of irradiation-induced winds is 
likely to be correct \citep{Maccarone04_metal}.
If neither qLMXBs nor MSPs show a strong metallicity dependence, then 
the idea of reduced magnetic braking in metal-poor GCs is correct \citep{Ivanova06_lmxb}.

\end{itemize}

\section*{Acknowledgments}

We thank Chris Deloye and Norman Murray for helpful discussions and an anonymous referee for
comments that helped to improve the paper.  
This work was supported by a Beatrice D. Tremaine Fellowship to NI. 
FAR was supported by NASA Grants NNG06GI62G and NNG04G176G.
JMF acknowledges support from Chandra theory
 grant TM6-7007X and Chandra Postdoctoral Fellowship
 Award PF7-80047.
KB acknowledges support from KBN grant 1P03D02228. 
COH acknowledges support from a Lindheimer Postdoctoral Fellowship, and from Craig Sarazin through Chandra grant G07-8078X.
 Simulations were performed on
CITA's Sunnyvale cluster, funded by the Canada Foundation 
for Innovation and the Ontario Research Fund for Research Infrastructure.

\bibliography{Ivanova}
\bibliographystyle{mn2e}

\end{document}